\documentclass[%
superscriptaddress,
preprint,
 amsmath,amssymb,
 aps,
]{revtex4-2}

\usepackage[a4paper, margin=0.7in]{geometry}

\usepackage{graphicx}% Include figure files
\usepackage{dcolumn}% Align table columns on decimal point
\usepackage{bm}% bold math
\usepackage{hyperref}% add hypertext capabilities
\usepackage{xcolor}

\usepackage{booktabs} 
\usepackage{multirow}
\usepackage{enumitem}
\usepackage{rotating}
\usepackage{xr}
\usepackage{nameref}

\newcommand{\QE}{{\sc Quantum ESPRESSO}}
\def\*#1{\mathbf{#1}}

\begin{document}

\title{Accurate and efficient protocols for high-throughput first-principles materials simulations}% Force line breaks with \\

\author{Gabriel de Miranda Nascimento}
\affiliation{Theory and Simulation of Materials (THEOS), École Polytechnique Fédérale de Lausanne, Lausanne 1015, Switzerland} 
\affiliation{PSI Center for Scientific Computing, Theory and Data, 5232 Villigen PSI, Switzerland} 
\affiliation{National Centre for Computational Design and Discovery of Novel Materials (MARVEL), Paul Scherrer Institute PSI, 5232 Villigen PSI, Switzerland} 

\author{Flaviano José dos Santos}
\author{Marnik Bercx}
\affiliation{PSI Center for Scientific Computing, Theory and Data, 5232 Villigen PSI, Switzerland} 
\affiliation{National Centre for Computational Design and Discovery of Novel Materials (MARVEL), Paul Scherrer Institute PSI, 5232 Villigen PSI, Switzerland} 

\author{Davide Grassano}
\affiliation{Theory and Simulation of Materials (THEOS), École Polytechnique Fédérale de Lausanne, Lausanne 1015, Switzerland} 

\author{Giovanni Pizzi}
\affiliation{PSI Center for Scientific Computing, Theory and Data, 5232 Villigen PSI, Switzerland} 
\affiliation{National Centre for Computational Design and Discovery of Novel Materials (MARVEL), Paul Scherrer Institute PSI, 5232 Villigen PSI, Switzerland} 

\author{Nicola Marzari}
\affiliation{Theory and Simulation of Materials (THEOS), École Polytechnique Fédérale de Lausanne, Lausanne 1015, Switzerland} 
\affiliation{PSI Center for Scientific Computing, Theory and Data, 5232 Villigen PSI, Switzerland} 
\affiliation{National Centre for Computational Design and Discovery of Novel Materials (MARVEL), Paul Scherrer Institute PSI, 5232 Villigen PSI, Switzerland} 

\date{\today}% It is always \today, today,

\begin{abstract}
Advancements in theoretical and algorithmic approaches, workflow engines, and an ever-increasing computational power have enabled a novel paradigm for materials discovery through first-principles high-throughput simulations.
A major challenge in these efforts is to automate the selection of parameters used by simulation codes to deliver numerical precision and computational efficiency.
Here, we propose a rigorous methodology to assess the quality of self-consistent DFT calculations with respect to smearing and $k$-point sampling across a wide range of crystalline materials. For this goal, we develop criteria to reliably estimate average errors on total energies, forces, and other properties as a function of the desired computational efficiency, while consistently controlling $k$-point sampling errors.
The present results provide automated protocols (named standard solid-state protocols or SSSP) for selecting optimized parameters based on different choices of precision and efficiency tradeoffs. These are available through open-source tools that range from interactive input generators for DFT codes to high-throughput workflows.
% on the required level of precision and accuracy.
%Throughout it, we strive to consistently cover all the concepts that one usually applies to the analysis of convergence of a specific system, implementing them in a systematic way that allows their extension for many materials and enables us to draw general conclusions for the construction of a protocol.
\end{abstract}

\maketitle

\section{Introduction}

Density-functional theory (DFT) has been a very successful approach to address the ground state of many-electron systems and has evolved into a predictive tool for materials discovery and design \cite{giustino2014materials, marzari2021electronic}.
%Given different possible implementations, this method is associated with a thriving ecosystem of computational codes, going from implementations of the Kohn-Sham equations using plane waves and pseudopotentials \cite{giannozzi2009quantum, giannozzi2017advanced, abinit, kresseEfficientIterative1996, kresseEfficiencyAbinitio1996}, to localized \cite{solerSIESTAMethod2002, g16} or mixed basis sets \cite{hutterCp2kAtomistic2014}, or solving the all-electron problem \cite{blaha2020wien2k}.
The predictive power of the theory, the robust implementations provided by DFT codes, and increasing available computational power in the past years have unlocked the potential of high-throughput approaches for materials discovery \cite{jainHighthroughputInfrastructure2011, mounetTwodimensionalMaterials2018b, merchant_scaling_2023}, where automated computational workflows are leveraged for the study and simulation of a large number of compounds. Starting from experimentally known crystal structures, these efforts are responsible for the construction and curation of numerous computational databases of materials \cite{jainCommentaryMaterials2013a, saalMaterialsDesign2013, curtaroloAFLOWAutomatic2012, haastrupComputational2D2018, Alexandria}, where several properties of many known compounds are addressed with various levels of accuracy. Moreover, data for these approaches serve as a starting point for strategies that aim at generating completely novel crystal structures, with approaches that range from random structure searching \cite{pickard2011ab}, structural prototyping \cite{haastrupComputational2D2018} to genetic algorithms \cite{zunger_genetic_algo, jennings2019genetic} and machine-learning-powered generative models \cite{xie2021crystal, zeni2023mattergen}.
% \todo{Perhaps we could make use of some more refs. here \red{GP: yes, in general in the intro. We should probably also cite e.g. random structure search, genetic algorithms, etc.}}

To properly manage these efforts on high-throughput calculations and support their scalability needs, workflow engines have been developed and coevolved with DFT codes.  They provide different levels of functionality that are suited for multiple use cases, ranging from input generation/output parsing and managing input/output operations, to interfacing with computational facilities and schedulers. Examples include materials APIs such as pymatgen \cite{pymatgen} and ASE \cite{larsenAtomicSimulation2017}, and workflow managers such as FireWorks \cite{jainFireWorksDynamic2015}, Pyiron \cite{Pyiron}, Atomic Simulation Recipes \cite{ASR} and AiiDA \cite{aiida, aiida1, UhrinEtAlWorkflowsAiiDAEngineering2021}.
% \red{GP: add citation of AiiDA engine}
%Notably, with a focus on the reproducibility of scientific results, engines such as AiiDA were also designed based on the concept of tracking the full provenance of data generated in these computational workflows, storing all results and relevant information in databases that are easily sharable and reproducible. 
In such frameworks, one of the main challenges in designing robust and efficient high-throughput workflows is achieving a reliable balance between precision and computational cost.
% Being quantitatively predictive is essential for leveraging these workflows for the discovery and screening of materials based on their properties, in a way that it is naturally important that they fully leverage the accuracy of the methods and codes that they run.
While carefully controlling the inner parameters of each method to achieve the most precise results for a given level of theory has been a standard practice in computational materials modeling since its origins, large-scale computational workflows also demand a stronger emphasis on computational efficiency, since they involve workloads that may scale to tens of thousands of materials and hundreds of thousands of DFT calculations. Optimizing the methods and inner parameters in these calculations such that they run as efficiently as possible, and ensuring that no computational resource is wasted and not converted into precision gains is, therefore, an essential part of designing robust computational workflows for first-principles materials simulations. Remarkable efforts in this direction include the creation of standard solid-state pseudopotential libraries (SSSP) \cite{sssp}, that established a testing protocol based on extensive DFT and density-functional perturbation theory (DFPT) calculations using the main pseudopotential libraries available \cite{PS1, PS2, PS3, PS4, PS5, PS6, PS7}.
Using these protocols, SSSP curates an exhaustive and open collection of extensively tested pseudopotentials and related parameters, suitable for different use cases of high-throughput and high-precision materials modeling.
%\red{GP: Should we mention in general pseudo libraries, and cite at least also the PseudoDojo here?}

This work contributes to solutions to these challenges by introducing the concept of standard solid-state protocols (SSSP) as an extended framework to the SSSP pseudopotentials \cite{sssp}. They are defined as a collection of parameters for specific codes and calculations, optimized for different levels of precision and computational efficiency and extensively tested for a wide range of materials. These operate within computational workflows, acting as an interface between their higher-level logic of operations and the inner-level details of parameters for a given calculation called by the workflow. More specifically, in this work we build upon the AiiDA workflow framework and its plugin and workflows for \QE{} \cite{giannozzi2009quantum,giannozzi2017advanced, giannozzi2020quantum, aiida-qe}, a plane-wave pseudopotential DFT code, but the results are general and applicable to any DFT code. In these protocols, we start from the results obtained by \citeauthor{sssp} \cite{sssp} for pseudopotentials and extend these to further benchmarks for the additional parameters that most strongly influence the precision and computational cost of DFT calculations for crystals: namely, $k$-point sampling and smearing techniques. For this reason, the SSSP acronym is extended to refer to protocols (and the pseudopotentials that they encompass). Similar work in this direction has been most notably performed by  \citeauthor{choudharyConvergenceMachine2019} \cite{choudharyConvergenceMachine2019}, where they employed a systematic procedure for benchmarking $k$-point grids and plane-wave energy cutoffs in calculations run with VASP \cite{kresseEfficientIterative1996, kresseEfficiencyAbinitio1996}, and has also been developed along the construction of materials databases, such as the Materials Project \cite{jainHighthroughputInfrastructure2011, jainCommentaryMaterials2013a}. However, we note that these works do not use smearing as a variable for optimization (employing a fixed value instead), missing out on the benefits that smearing techniques may bring to the convergence of Brillouin zone sampling, either by reducing the required computational cost for the calculations or improving their precision. In our work, we simultaneously consider $k$-point sampling and smearing temperature in the optimization process and combine these results with the guidelines from \citeauthor{sssp} \cite{sssp} to create reliable sets of parameters that are suited for different tradeoffs of precision 
%\red{GP: In general, we should probably never speak of accuracy but of precision?}
and computational time. These parameters are then extensively tested for a wide distribution of materials.
% In these benchmarks, as previous works, we employ uniformly distributed $k$-point meshes for $k$-point sampling, which are the \textit{de facto} standard in high-throughput DFT calculations, noting that other advanced methods for Brillouin zone integration exist.
 
In this work, we start by summarizing the general interplay between $k$-point sampling and smearing techniques targeted at improving the convergence and efficiency of DFT calculations. We then describe the automated workflows developed for benchmarking the choice of parameters and their effect on relevant physical properties. We identify distinct convergence behavior for various classes of materials and establish general criteria for which optimized parameters can be found. Finally, we discuss how these results are integrated in standard solid-state protocols, showcasing the different ways they are available for users through the open-source ecosystem of these codes.

\section{Smearing and $k$-point convergence}
One of the main keys to numerical precision in all-electron and pseudopotential DFT calculations for crystals is $k$-point sampling. In this framework, errors arise due to the discretization of the Brillouin zone (BZ) used to evaluate integrals of $k$-dependent functions, most notably the total energy \cite{giustino2014materials, dal1996pseudopotential}. Errors in total energies can also propagate to other quantities; e.g., forces on atoms and stresses.
%through the Hellmann–Feynman theorem \cite{hellmann1933rolle, feynman1939forces}.
Numerical issues become even more evident for metallic systems, whose occupation function $f(\*k)$ becomes discontinuous at the Fermi surface. Integration of these functions has notorious poor convergence with respect to a uniform sampling mesh \cite{methfesselHighprecisionSampling1989} since they are not differentiable. Practically, performing quantitatively accurate calculations for metals would, in this case, demand an exceedingly high number of $k$-points, making the calculations computationally unfeasible. %, as the cost of a DFT calculation scales linearly with the number of $k$-points.  

To solve this problem, various smearing techniques were developed to effectively smooth out the discontinuity of the electronic occupations at the Fermi level (see Ref. \cite{dossantosFermiEnergy2023} for a summary and review). By replacing the discontinuous occupation function with a smooth and differentiable one, smearing techniques are able to achieve exponential convergence of integrals with respect to the number of $k$-points. This process is often physically interpreted as adding a fictitious electronic temperature ($\sigma$) to the system \cite{de1995lattice}, since the broadening of the electronic occupations gives rise to a smeared density of states that can be written as the addition of an entropic term to the total free energy.  By modifying $f(\epsilon)$ from the original physical one at $T=0$K, the integrals converge faster, but the price to be paid is that they converge to a different value than the $\sigma\rightarrow0$ limit. More specifically, the smearing entropy that now enters the generalized free energy can be written as a power expansion in the smearing temperature \cite{dossantosFermiEnergy2023}:

\begin{equation}
\label{eq:entropy}
    S=\sum_{k=0}^{\infty} c_k n^{(k-1)}(\mu) \sigma^k,
\end{equation}
with coefficients $c_k$ of the form:

\begin{equation*}
    c_k=(-1)^{k+1} \frac{1}{k!} \int_{-\infty}^{\infty} \epsilon^{k+1} \tilde{\delta}(\epsilon) d \epsilon,
\end{equation*}
where $\tilde{\delta}$ is the broadening function employed.

Eq. (\ref{eq:entropy}) leads us to a general observation that $S$ depends not only on the smearing temperature $\sigma$, but also on the value of the derivatives of the density of states at the Fermi energy $n^{(k-1)}(\mu)$. As further discussed, these two factors play an important role in driving the deviations from the $\sigma=0$ total energy when introducing smearing.

\begin{figure}[!h]
    \centering
    \includegraphics[width=\linewidth]{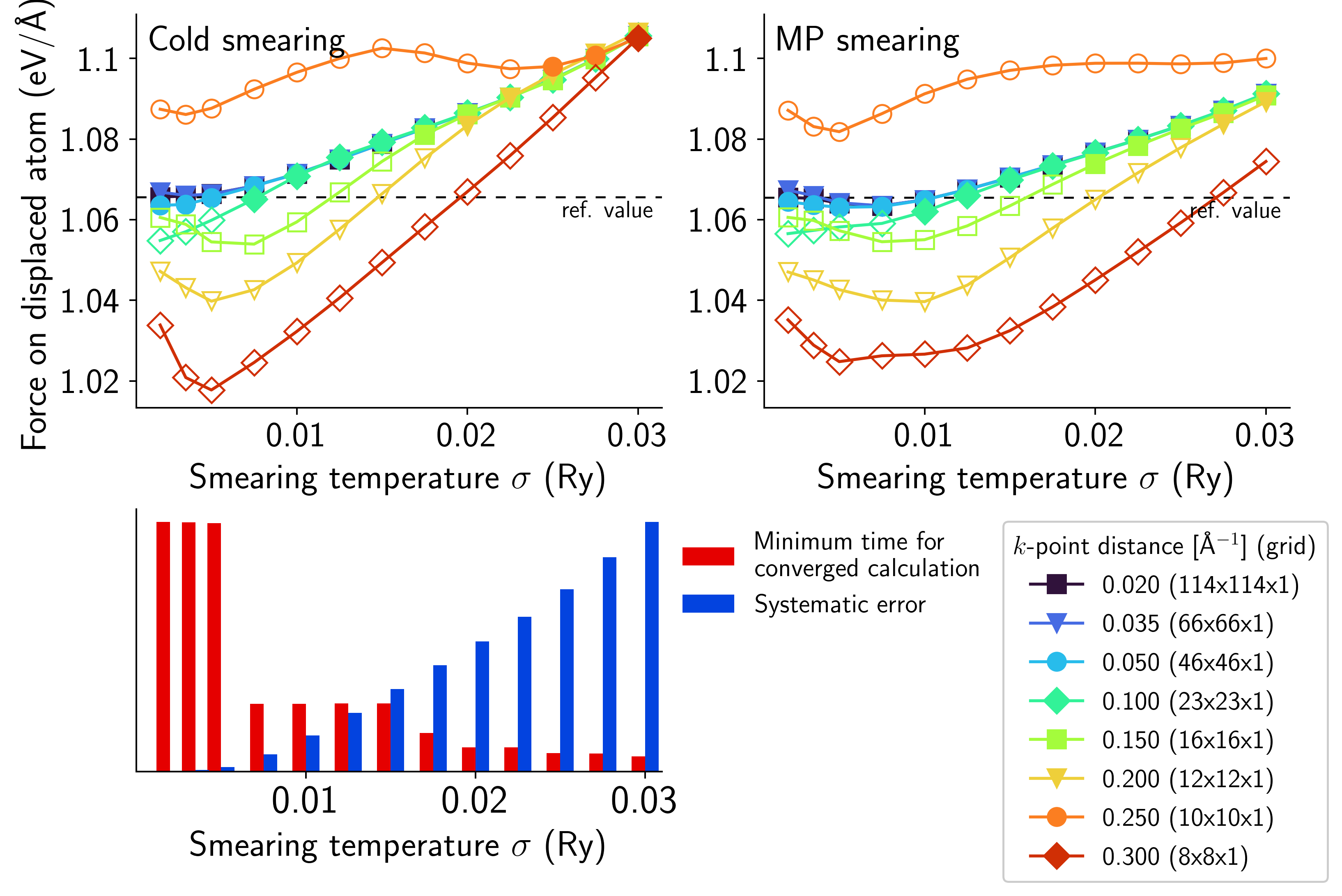}
    \caption{\textbf{Convergence of forces as a function of $k$-point mesh and smearing for the metallic 1T phase of two-dimensional MoS$_2$}. Top: DFT-calculated force on an atom slightly displaced away from its equilibrium position as a function of smearing temperature, using Marzari-Vanderbilt cold smearing \cite{marzari1999thermal} (left) and Methfessel-Paxton first-order smearing \cite{methfesselHighprecisionSampling1989} (right), for distinct $k$-point meshes. The horizontal dashed line indicates a reference value taken at the limit of the lowest smearing temperature (0.002 Ry), for which the 3 densest $k$-point meshes agree (up to 0.0035 eV/\AA). 
    % Markers in this plot are filled based on whether the values agree with those at denser $k$-point meshes, up to this same tolerance
    Data points in the plot are shown as filled symbols if their calculated values match the results obtained using the densest k-point meshes within a tolerance of 0.0035 eV/Å. Empty symbols indicate that the results differ by more than this tolerance.
    % \red{GP: previous sentence unclear?} \GM{changed \checkmark}.
    Bottom: minimum amount of computational wall time required to converge a calculation w.r.t. $k$-points (red bars), for fixed threshold of 0.0035 eV/\AA, and systematic error introduced (blue bars), both as a function of the smearing temperature $\sigma$, for the calculations done with cold smearing (upper left plot).
    \label{fig:MoS2}}
\end{figure}

The practical interplay between $k$-point sampling and smearing temperature can be understood by tracking the value of a DFT-calculated property (e.g. total energy or forces) as a function of different combinations of parameters. This is done as an example in Figure \ref{fig:MoS2} for the metallic 1T phase of two-dimensional MoS$_2$. In the upper plots in the figure, we show the calculated value of the force on an atom, slightly displaced out of its equilibrium position, as a function of the smearing temperature $\sigma$ and $k$-point sampling for two smearing methods: Marzari-Vanderbilt cold smearing \cite{marzari1999thermal} and first-order Methfessel-Paxton \cite{methfesselHighprecisionSampling1989} (see Ref. \cite{dossantosFermiEnergy2023} for a review of convergence behavior according to different smearing techniques).
% \blue{FJdS what is the smearing type? It might be useful to mention it here and in the figure caption}.
From these results, we observe distinct behavior in the two limits of $\sigma$. In the $\sigma \rightarrow 0$ limit, the aforementioned problem of poor $k$-point sampling convergence becomes evident. Due to the fact that no variational principle exists for the convergence of the total energy (and other properties) with respect to $k$-sampling, results obtained by various coarse $k$-point grids are essentially random. That is, a small change in $k$-point sampling in this regime may lead to large changes in the calculated property. In this work, we understand that this source of errors due to finite sampling is dangerous and hard to quantify and should always be avoided; we call it the statistical (sampling) error. In this low-smearing regime, satisfiable convergence is only achieved for extremely dense $k$-point grids (with mesh sizes approximately above $50\times50$
% \todo{mention the choice of 2D and/or 3D?}
in the case of this 2D material), which is impractical for high-throughput workflows and other computational approaches.

% \begin{figure}[!h]
%     \centering
%     \includegraphics[width=\linewidth]{figures/fig1.png}
%     \caption{\textbf{Convergence of forces as a function of $k$-point mesh and smearing for the metallic phase of two-dimensional MoS$_2$}. Top: DFT calculated force on an atom randomly displaced outside its equilibrium position as a function of smearing temperature (as prescribed by Marzari-Vanderbilt cold smearing \cite{marzari1999thermal}), for distinct $k$-point meshes. The horizontal dashed line indicates a reference value taken at the limit of the lowest smearing temperature (0.002 Ry), for which the 3 densest $k$-point meshes agree (up to 0.003 eV/\AA). 
%     % Markers in this plot are filled based on whether the values agree with those at denser $k$-point meshes, up to this same tolerance
%     Data points in the plot are shown as filled symbols if their calculated values match the results obtained using the densest k-point meshes within a tolerance of 0.003 eV/Å. Empty symbols indicate that the results differ by more than this tolerance.
%     % \red{GP: previous sentence unclear?} \GM{changed \checkmark}.
%     The inset is the crystal structure of 2D MoS$_2$. Bottom: minimum amount of computational wall time required to converge a calculation w.r.t. $k$-points (red bars), for fixed threshold of 0.003 eV/\AA, and systematic error introduced (blue bars), both as a function of the smearing temperature $\sigma$.
%     \label{fig:MoS2}}
% \end{figure}

On the other hand, we observe an enhanced $k$-point convergence as the smearing temperature increases. The higher $\sigma$ is, the coarser the $k$-point meshes that can reproduce the results from the very dense $k$-sampling limit. This enables more computationally efficient calculations to be run while suppressing the source of errors that arise from poor $k$-point sampling. However, the problem with large $\sigma$ is that higher entropic contributions kick in, so that the values for energies and other calculated properties start deviating from the true $\sigma \rightarrow 0$ limit. This is observed in the upper plot in Figure \ref{fig:MoS2}, where the curves for the different $k$-point meshes converge to the same values for high smearing, but deviate from the reference converged value $\sigma \rightarrow 0$ (calculated with the densest $k$-point mesh and the smallest $\sigma$). In the following, we will refer to the errors that arise from smearing alone (i.e. when $k$-point sampling is properly converged) as systematic (smearing) errors. Differently from sampling errors, systematic errors are better behaved, i.e., small changes in $\sigma$ lead to small variations of the corresponding errors.

The optimized choice of $k$-point sampling and smearing parameters must be such that the interplay between systematic and sampling errors is kept under control.
This ``sweet spot", at which smearing effectively suppresses sampling errors but does not generate excessive systematic errors, is naturally dependent on the compromises between precision and computational cost one is willing to make.
An effective control of errors should therefore provide a reliable way of setting this tradeoff, making sure that increments in computational cost are effectively converted into precision and vice-versa.
In this work, we use smearing as a reliable parameter for suppressing the statistical errors introduced by $k$-point sampling, while using systematic errors as a controllable variable to adjust the tradeoff between precision and computational cost.
The generalization of convergence studies as the one shown in Figure \ref{fig:MoS2} to a wide and representative range of materials gives rise to our defined SSSP protocols. They form the set of statistically relevant parameters optimized for different precision vs. cost tradeoffs, while also being able to account for relevant differences in electronic structure across the materials landscape and thus providing an efficient framework for high-throughput workflows.

\section{Results}
\subsection{The benchmarking workflows}
\label{sec:workflow_details}

The optimization of parameters relies on data from DFT calculations from a wide variety of materials and combinations of parameters. As discussed in the Methods section, 269 structures were sampled from the Materials Cloud databases \cite{mounetTwodimensionalMaterials2018b, campiExpansionMaterialsCloud2023, campi2022mc2d, huber2022mc3d} of three-dimensional (MC3D) and two-dimensional (MC2D) structures and used as starting point of the workflows discussed in this work. For each structure, an atom in the unit cell is randomly selected and displaced towards its nearest neighbor by 5\% of the corresponding bond length. This avoids high-symmetry configurations where forces vanish by symmetry. Next, various DFT self-consistent calculations are performed with different combinations of $k$-point sampling and smearing parameters. We define a variable $\lambda$ to represent the sampling of the BZ in each calculation. With units of $1/\textrm{\AA}$, $\lambda$ encodes the maximum distance in reciprocal space, along each of the three directions defined by the reciprocal lattice vectors, between adjacent $k$-points of a uniform $\Gamma$-centered $k$-point mesh used in the calculation. 

For smearing, Marzari-Vanderbilt cold smearing \cite{marzari1999thermal} is chosen for all calculations, with $\sigma$ ranging from 0.002 to 0.03 Ry.
For insulators, we also perform calculations with fixed occupations and no smearing broadening, that can be used as $\sigma = 0$ reference value in this case. 
The choice of the cold smearing method is based on the fact that this broadening function is devised to yield total free energies independent of the smearing temperature up to and including the second order (from the entropy expansion in Eq. \ref{eq:entropy}), excluding the need of \emph{a posteriori} corrections that are otherwise needed, for instance, in the Fermi-Dirac and Gaussian smearing techniques \cite{dossantosFermiEnergy2023}. Furthermore, $\tilde{\delta}$ in this approach produces a positive-definite occupation function, unlike the approach from \citeauthor{methfesselHighprecisionSampling1989} \cite{methfesselHighprecisionSampling1989},  where occupations can be unphysically negative.
Nonetheless, we note that the same approach for benchmarking DFT quantities and statistical considerations used in the analysis of errors for the creation of protocols in this work are compatible with the other aforementioned smearing methods, readily available in multiple DFT codes, although the numerical values for optimized parameters may differ. For instance, we expect that repeating this benchmark for Gaussian smearing will result in smaller optimized values for the smearing temperature, given its quadratic depedence of the free energy with respect to $\sigma$, while requiring finer $k$-point meshes for equivalent precision (optimized values for $\sigma$ using the Fermi-Dirac distribution can be obtained by scaling the $\sigma$ parameters for Gaussian smearing down by a factor of $\lambda=\sqrt{2/3}\pi$, as pointed out in Ref. \cite{dossantosFermiEnergy2023}). On the other hand, the numerical values obtained by our benchmarks using cold smearing can be used as a first approximation for resonable calcuations with other smearing techniques with similar $\sigma$ dependence of the free-energy. This is the case of first-order Methfessel-Paxton. As illustrated in Figure \ref{fig:MoS2}, the dependence of DFT energies and forces as a function of $\sigma$ in the range considered here is in most cases similar enough between the two smearing techniques, suggesting that converged parameters for cold smearing are likely to be appropriate choices also for Methfessel-Paxton smearing.

%In this way, we note that the numerical values obtained by our protocols are best applied to the Fermi-Dirac method by rescaling $\sigma$ by a factor of $(\sqrt{2/3}\pi)^{-1}$, as prescribed by ref.\cite{dossantosFermiEnergy2023}\red{GP: this is not true, I think! The factor is between Gaussian and FD, but our values are for cold - if we want to say this, we can say that if we do this for e..g Gaussian, then one does not need to do FD} \blue{FJdS: I agree}, and that the protocols here devised do not exempt the need for \emph{a posteriori} corrections of the quadratic dependence of the free energy with $\sigma$ for the Gaussian and Fermi-Dirac cases. 

As shown in Supplementary Table S1, 
%\red{GP: I would call these supplementary Table S1, Suppleentary Figure S4, ... - i.e. use arabic numbers and prepend a letter S},
this process generates 91 distinct calculations for each metallic structure and 28 for each insulating one, that can be labeled by the corresponding pair $(\lambda, \sigma)$. The workflow automatically checks for electronic convergence (achieved in all cases) and keeps track of the final results for total energies ($E$), the force on the displaced atom ($F$), and pseudo stresses $\Sigma$,
%\red{GP: shall we call this "pseudo stress" and indeed explain just after, saying "defined as" rather than "more precisely"?}
defined as the average of the diagonal components $\sigma_{ii}$ of the stress tensor, where $i=x,y,z$ for 3D materials and $i=x,y$ for 2D (we use $\Sigma$ in this case, given the customary use of $\sigma$ in this context for smearing temperature).

\subsection{Analysis of errors}
\label{sec:analysis_of_errors}

For each structure, a specific set of parameters must be chosen as the reference point to compare all other calculations. Ideally, this reference should be taken in the limit $\lambda \rightarrow 0$ and $\sigma \rightarrow 0$, i.e., an infinitely dense $k$-point mesh and zero smearing temperature, which is however not practically attainable. Therefore, for $k$-point sampling, we always take as a reference the densest mesh $\lambda_{\textrm{ref}}=0.035 \textrm{ \AA}^{-1}$. This value of $\lambda$ provides $k$-point meshes with, on average, 202 and 888 irreducible $k$-points per atom in the primitive cells for 2D and 3D materials, respectively. The choice $\sigma_{\textrm{ref}}$ for each material should instead guarantee that results are properly converged with respect to $k$-points. Therefore, $\sigma_{\textrm{ref}}$ is chosen for each structure as the lowest value of $\sigma$ for which the calculations with the three densest $k$-point meshes ($\lambda = 0.035, 0.05, 0.10 \; \textrm{\AA}^{-1}$) have total energies, forces, and (pseudo) stresses converged on $k$-points below given tolerances $\tau^k_P$, with $P\in\{E, F, \Sigma\}$. This means that sampling errors introduced on total energies, forces, and stresses, defined respectively as $\varepsilon^k_E$, $\varepsilon^k_F$, and $\varepsilon^k_\Sigma$, are such that:
\begin{equation}
\label{eq:kpoints_error_thr}
        \varepsilon^k_P \leq \tau^k_P, \quad \textrm{where} \quad \varepsilon^k_P(\lambda, \sigma) \equiv |P_{(\lambda, \sigma)} - P_{(\lambda_{\textrm{ref}}, \sigma)}|,
\end{equation}
where $P_{(\lambda, \sigma)}$ is the numerical value of the property $P\in\{E, F, \Sigma\}$ computed in the DFT calculation having $(\lambda, \sigma)$ parameters.

These $\tau^k_P$ thresholds represent maximum limits for the tolerable amount of sampling errors in a calculation and should be analyzed separately from the systematic errors introduced by smearing, $\varepsilon^s_P$. The latter is defined as the difference in the property value computed between the calculation with smearing $\sigma$ and $k$-point distance $\lambda_{\textrm{ref}}$ and the reference calculation with ($\lambda_{\textrm{ref}}, \sigma_{\textrm{ref}}$):
\begin{equation}
        \varepsilon^s_P(\sigma) \equiv |P_{(\lambda_{\textrm{ref}}, \sigma)} - P_{(\lambda_{\textrm{ref}}, \sigma_{\textrm{ref}})}|.
\end{equation}

In this sense, sampling errors and systematic errors, $\varepsilon^k_P(\lambda, \sigma)$ and $\varepsilon^s_P(\sigma)$ respectively, are calculated differently, as the former considers comparisons within the same smearing value and different $k$-point meshes, while the latter between the same (reference) $k$-point mesh and different smearing temperatures. The sum of the two errors is not simply the the total error in a calculation $(\lambda, \sigma)$, but is an upper bound to it: 
\begin{equation}
        \varepsilon^{\textrm{Total}}_P(\lambda, \sigma) \leq \varepsilon^k_P(\lambda, \sigma) + \varepsilon^s_P(\sigma),
\end{equation}
as both errors can be either positive or negative.

Analyzing these two sources of errors separately, thus, eliminates misleading cases where sampling and systematic errors cancel out by chance. Additionally, defining numerical values for the convergence thresholds $\tau^{k, s}_P$ a priori is an ill-posed task, as one is ultimately interested in their propagating effect on errors of measurable properties of materials. Establishing a robust heuristic for determining these thresholds and using them to reliably control the precision vs. cost tradeoffs in the calculations is one of the purposes of the analysis that follows.

\subsection{General convergence trends}

\begin{figure}[hbtp]
    \centering
    \includegraphics[width=\textwidth]{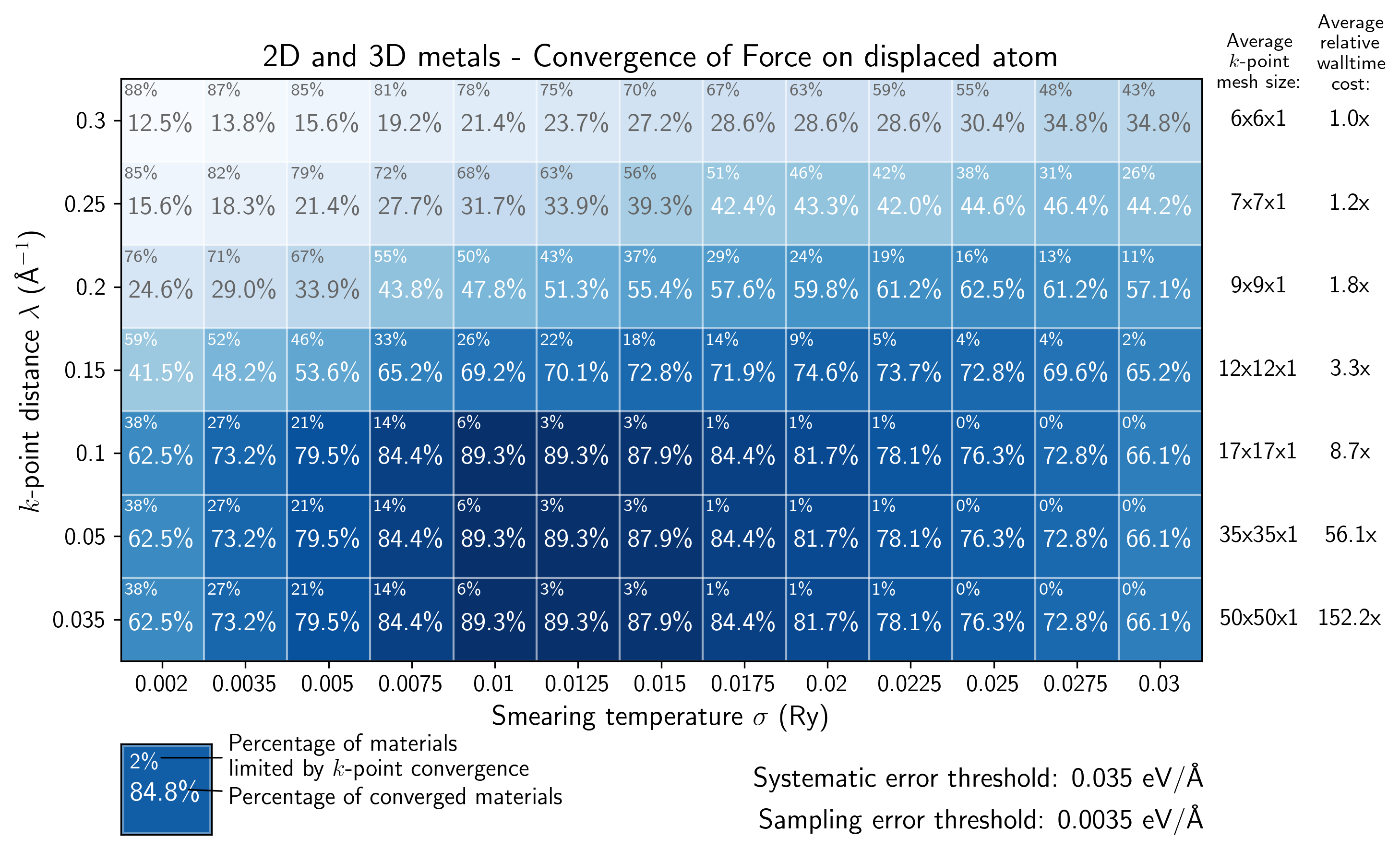}
    \caption{\textbf{Force convergence for all 2D and 3D metals as a function of the $k$-points distance and smearing temperature}. ``Convergence", in this case, is considered reached by calculations producing both systematic and sampling errors below the thresholds $\tau^s_F$ and $\tau^k_F$, respectively. We use $\tau^s_F = 0.035$ eV/{\AA} and $\tau^k_F = 0.0035$ eV/{\AA} for this plot, noting that similar qualitative results are obtained with different threshold values. A visual legend is presented for an additional guide on the other metrics displayed in this plot. The metric on the relative wall time cost is defined as the ratio between the wall time required to run calculations for all materials with the respective $k$-point distance parameters and the wall time of the corresponding calculations with the computationally least expensive parameters in the analysis (least dense $k$-point meshes).
    % \todo{Not really sure if Giovanni meant landscape mode when he referred to ``make it full page".}\red{GP: No, I don't think so. Just avoid any white margin on left/right, put it back in portrait mode.}
    }
    \label{fig:2D-3D-heatmap}
\end{figure}

A first observation of the convergence trends for the calculated structures is shown in Figure \ref{fig:2D-3D-heatmap}, for the specific case of metals, where we plot the percentage of materials whose calculated forces are converged as a function of the calculation parameters. Similar plots for errors in total energies and stresses have qualitatively similar characteristics as the one shown for forces. Here, both error contributions arising from $k$-points convergence and smearing are taken into account, such that a calculation is considered ``converged" if and only if $\varepsilon^k_F \leq \tau^k_F$ and $\varepsilon^s_F \leq \tau^s_F$. Although the values for $\tau_F$ are still free parameters in the analysis, qualitative conclusions regarding the convergence behavior of the calculations can be drawn. First, the poor convergence at very low $\sigma$ values can be observed on the left part of Fig. \ref{fig:2D-3D-heatmap}, even at very dense $k$-point meshes. Indeed, we can precisely quantify the percentage of materials that are bounded by $k$-point errors (as opposed to being dominated by systematic errors) by taking the limit of $\tau^s_F \rightarrow \infty$ at a fixed $\tau^k_F$. This is plotted at the upper left
%\red{GP: Upper left?}
corner of each square in the heatmap of Fig. \ref{fig:2D-3D-heatmap}. We can see that calculations for low values of $\sigma$ are mostly not converged due to sampling errors, and this metric monotonically decreases with $\sigma$, i.e., increasing the smearing temperature effectively enhances $k$-point convergence for all materials, as expected. However, total convergence also decreases on the right extreme of Fig. \ref{fig:2D-3D-heatmap}.
%for calculations with $\sigma > 0.0225$ Ry\red{GP: why this specific value? I see a decrease even before}.
In this regime, instead, the errors in the calculations are driven by the systematic errors introduced by the high smearing temperature. This highlights that an intermediate value of $\sigma$ exists that minimizes both sampling and systematic errors for a fixed $k$-point mesh.

Furthermore, we can also note another trend when following the convergence vertically across $k$-point meshes in Fig. \ref{fig:2D-3D-heatmap}. As the density of $k$-points decreases ($\lambda$ increases), the calculations become less computationally demanding, marked by the decrease in the relative wall time cost (numbers at the bottom of each square). Total convergence also decreases, but importantly we observe that the ``sweet spot" in smearing also shifts to the right as $\lambda$ increases. This produces a diagonal trend on the heatmap of Fig. \ref{fig:2D-3D-heatmap}, highlighting the fact that the ``coarser" (i.e. less dense) a $k$-point mesh is, the more it benefits from a higher smearing temperature. This concept is at the core the optimization in protocols for high-throughput calculations: leveraging smearing temperature to enable more computationally efficient calculations with robust control of errors. 

\subsection{Materials with open $f$-shell elements}
\label{sec:lanthanides}

\begin{figure}[t]
    \centering
    \includegraphics[width=\linewidth]{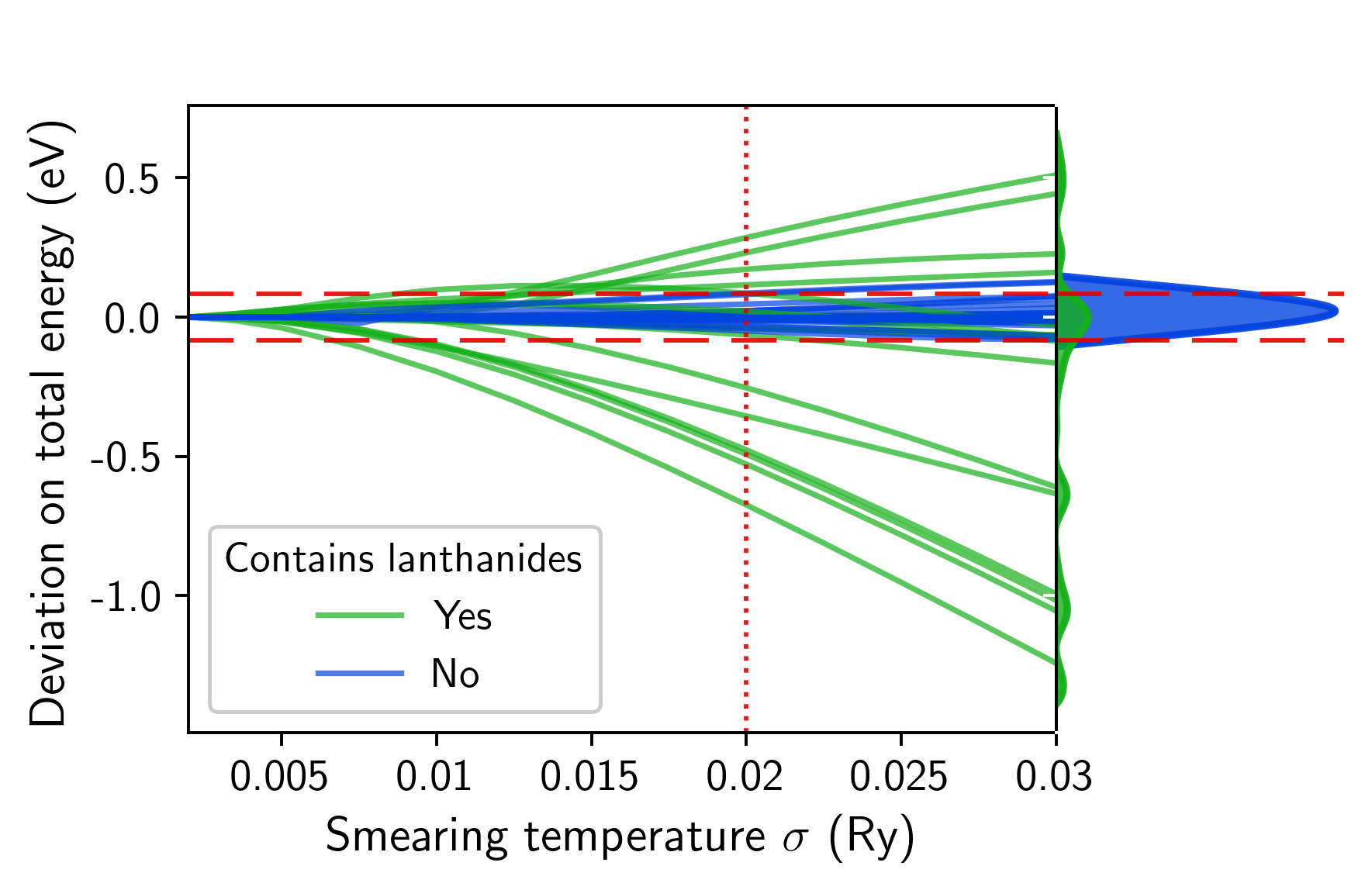}
    \caption{\textbf{Systematic error on total energies as a function of smearing temperature}. Materials with lanthanides are indicated by green lines and are those most sensitive to systematic errors. The curves on the right represent the distribution of errors at $\sigma = 0.3$ Ry for each class of materials (values interpolated from a discrete histogram). Horizontal red dashed lines indicate the error margins within which 85\% of materials lie ($\pm$0.082 eV), measured at an intermediate smearing value of 0.2 Ry (dotted vertical line). Only calculations with the densest $k$-point meshes are considered, known to be converged with respect to $k$-points.}
    \label{fig:systematic_errors_lanthanides}
\end{figure}

From our calculations, metallic materials containing lanthanide elements form the group of compounds most strongly affected by errors in total energies, arising mainly from the smearing temperature. This is illustrated in Figure \ref{fig:systematic_errors_lanthanides}, where we plot the deviation on the total energy as a function of $\sigma$ for each material. In the limit of high $\sigma$, materials containing lanthanides have remarkably higher systematic errors than the rest of the materials that do not contain open $f$-shell elements. These errors stem from the electronic structure of these materials, where valence $f$-states, due to their high localization in real space, generate low $k$-dispersive states in the system band structures, contributing to sharp peaks in the density of states. If these bands are close to the Fermi energy, these peaks lead to high derivative values $n^{(k-1)}(\mu)$, which in turn lead to larger $S$ with increasing $\sigma$ (see Eq. \ref{eq:entropy}), resulting in high deviations of total energies observed in Figure \ref{fig:systematic_errors_lanthanides}.
In fact, the pathological case of lanthanide materials with smeared occupations was also observed by \citeauthor{bosoni_how_2023} \cite{bosoni_how_2023}, relating the strong deviation of the total free energy with smearing temperature with the presence of narrow $f$-bands close to the Fermi level.
We also generate evidence for this hypothesis by calculating the projected density of states (PDOS) for all lanthanide-containing materials from Fig. \ref{fig:systematic_errors_lanthanides}, see Supplementary Figure S1. All lanthanide materials that have large systematic errors have indeed their electronic DOS near the Fermi level dominated by sharp peaks arising from the lanthanide $f$-states. Therefore, structures containing lanthanide or actinide elements form a class of materials for which protocols should take special care due to their higher sensitivity to systematic errors introduced by increasing $\sigma$.

% As a note, we stress that all PDOS calculations were performed with the protocols specified in Section \ref{sec:methods_workflows}, employing the same GGA xc-functional. We note that more advanced techniques, such as Hubbard corrections in the DFT+U framework \cite{haddadi2024site, timrov2018hubbard}, may break some degeneracies in the partially filled $f$-manifolds in these cases, opening an electronic gap where these convergence errors are less of a problem. However, as these protocols are designed for standard DFT-GGA calculations, we understand that they should still take these problems into consideration to deliver the highest fidelity results within this level of theory. 

\subsection{Multiobjective optimization of precision and wall time}

\begin{figure}
    \centering
    \includegraphics[width=0.8\linewidth]{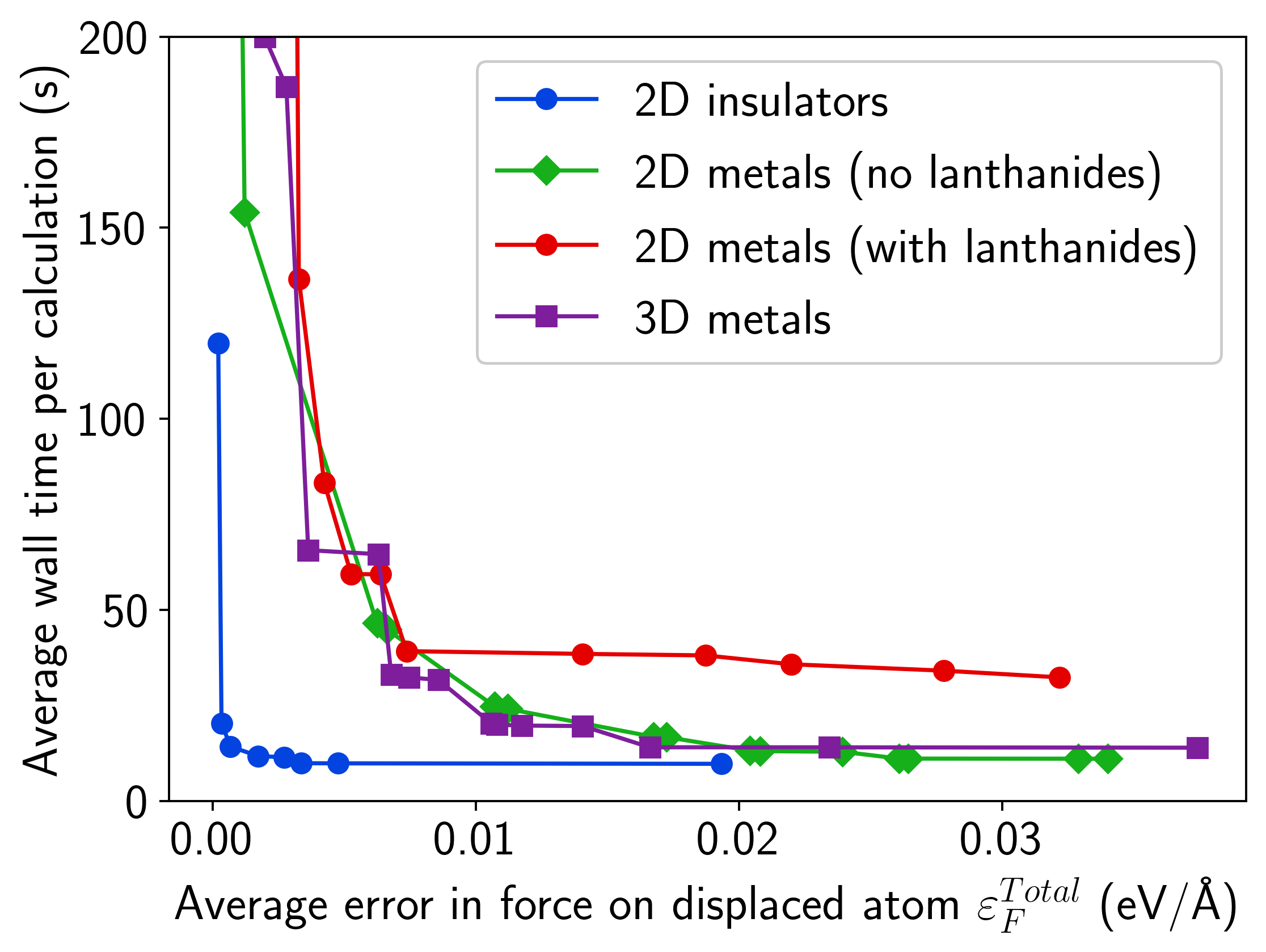}
    \caption{
    %\red{GP: Move 2D insulators as the very first one in the legend.}
    \textbf{Pareto front between average wall time per calculation vs. average total errors on forces on the displaced atom ($\varepsilon^{\textrm{Total}}_F$)}. Curves for the different material classes considered in this work are shown.
    Materials with lanthanides are defined as having at least one lanthanide element in their composition. All 3D metals considered here do not contain lanthanides.
    % \red{GP: clarify that 2D metals is all *without lanthanides", and (lanthanides) indicated that at least a lanthanide element is included (right?)}
    }
    \label{fig:Pareto-front-all-classes}
\end{figure}

% Besides the robust control of errors and the classes of materials most susceptible to them, protocols focus by design on the central problem of maximizing accuracy while minimizing computational time. In this way, 
Only a subset of all parameter combinations tested in our benchmarks minimize computational wall time for a fixed amount of average errors. This Pareto front, illustrated in Figure \ref{fig:Pareto-front-all-classes}, represents a collection of parameters that are most effective in converting increased computational resources into precision and vice-versa.
Comparing the curves for 2D insulators and 2D metals, the faster convergence behavior of insulators is clear. As expected, these systems do not have the discontinuity problem at the Fermi level as metals do and, therefore, require less dense $k$-point meshes to achieve the same level of errors on calculated forces. This highlights opportunities for protocols in saving resources for calculations in systems known to be insulating. On the other hand, the curve of 2D metals including lanthanides shows that the higher systematic errors introduced by smearing for materials with open $f$-shells impose the need for higher computational resources to achieve the same level of precision. Furthermore, by comparing 2D and 3D metals, we do not observe a considerable difference in behavior with respect to the parameters that minimize the computational wall time for a given level of precision in the calculations. As such, in the following we group these two groups of materials into a single class of metals.

The analysis from Fig. \ref{fig:Pareto-front-all-classes}, however, deals with sampling and systematic errors on equal footing. Nevertheless, as discussed earlier, using smearing temperature to suppress the less controllable sampling errors is beneficial, since it provides a more controllable source of errors according to desired precision $\times$ cost tradeoff. A reliable Pareto front for our optimized parameters, in this way, takes that into consideration by imposing additional constraints on the composition of the total error in the calculations. In this work, we impose the constraint that, within any calculation, tolerable sampling errors should be at most 5\% of the value of systematic errors. This ensures that, at any level of precision required, smearing correctly guarantees the convergence of $k$-point sampling, preventing that any slight change in the materials electronic structure or in the sampling of the Brillouin zone generates uncontrollable errors in the calculations. 

\begin{figure}
    \centering
    \includegraphics[width=\linewidth]{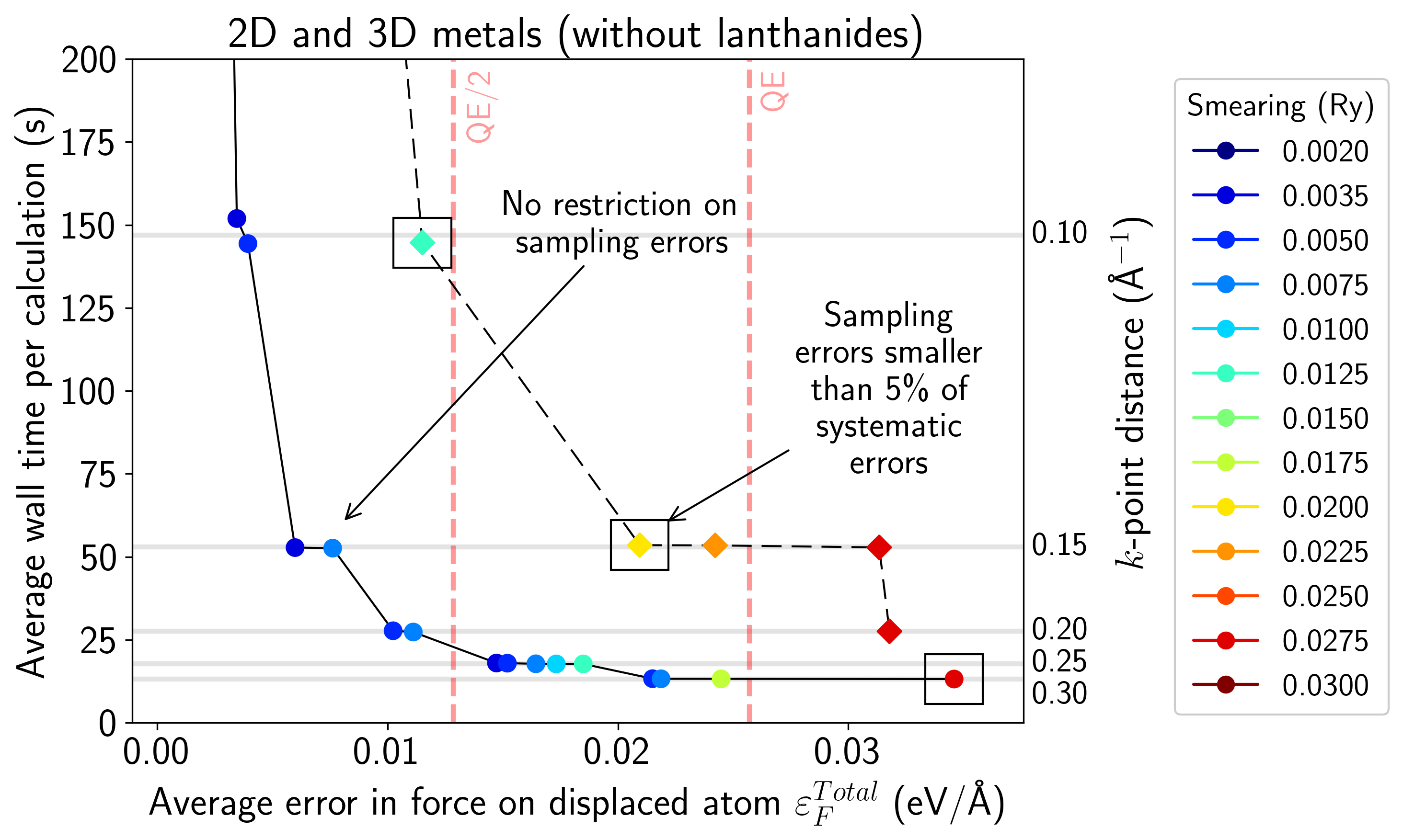}
    \caption{\textbf{Optimized parameters on the Pareto front between precision and computational cost.} The $x$-axis represents the average total error on the force on the displaced atom ($\varepsilon^{\textrm{Total}}_F$) for all metals without lanthanides (2D and 3D). The $y$-axis represents the average computational wall time per calculation, which has a direct mapping to the $k$-point mesh size (axis on the right). The dashed black line represents the Pareto front with the additional constraint of sampling errors being smaller than 5\% of the systematic errors ($\varepsilon^k_F < 0.05 \varepsilon^s_F$), while the continuous line is the unconstrained one. The vertical red dashed lines represent reference thresholds for forces as the stopping criteria in a DFT relaxation. The line on the right (\texttt{QE}) represents the default value of {\sc Quantum ESPRESSO} of $10^{-3}$ Ry/bohr (0.026 eV/{\AA}), while the one on the left (\texttt{QE/2}) represents a stricter reference of $5\cdot10^{-4}$ Ry/bohr (0.013 eV/{\AA}, i.e. half of the \texttt{QE} value). Optimized parameters are chosen on this Pareto front and indicated by black rectangles (see Discussion section). As each data point is colored according to the value of $\sigma$, the optimized parameters $(\lambda, \sigma)$ can be directly read out of the plot.}
    \label{fig:pareto_front_constrained}
\end{figure}

Figure \ref{fig:pareto_front_constrained} illustrates the analysis of the Pareto front for metals with the additional constraint on error composition, compared to the unconstrained version shown in Figure \ref{fig:Pareto-front-all-classes}. By considering sampling and systematic errors distinctly, we observe that the additional constraint shifts the suggested $k$-point sampling values to denser meshes (i.e. smaller $\lambda$) in order to minimize sampling errors in favor of systematic ones. In doing so, the suggested values for the smearing temperature increase accordingly. Furthermore, in order to quantify the magnitude of errors, we can compare errors with the thresholds on forces used in relaxation calculations. Having forces on all atoms below a threshold is one of the criteria considered for stopping the ionic minimization and atomic position updates in relaxation calculations. Specifically, in Fig. \ref{fig:pareto_front_constrained} we show the default forces threshold employed in {\sc Quantum ESPRESSO}, together with a stricter reference of half of this value.

An additional and important advantage of the analysis presented in Fig. \ref{fig:pareto_front_constrained} over the convergence heatmaps of Figure \ref{fig:2D-3D-heatmap} is its higher independence of the thresholds considered. First, at variance with the heatmap analysis, the Pareto front approach is totally free from a threshold on systematic errors ($\tau^s$), as the average error in the calculations is presented as an independent variable on the $x$-axis (as opposed to serving as a fixed cutoff value as in Fig. \ref{fig:2D-3D-heatmap}). The only adjustable threshold parameter, in this case, becomes $\tau^k$, as it is needed to define the lowest value of smearing producing reliable calculations for a material ($\sigma_\textrm{ref}$), according to the definition of Eq. \ref{eq:kpoints_error_thr}. However, we also observe that, in practice, the optimized parameters obtained from these analyses are independent of $\tau^k$ over a wide range of values, as illustrated in Supplementary Figure S2. This can be attributed to the design choice of suppressing sampling errors in favor of systematic errors, contributing to an additional level of robustness in the analysis.

Finally, as illustrated in Supplementary Figure S3, the optimized parameters that minimize the computational time and errors for energies or stresses are effectively the same as those for forces presented here. Therefore benchmarking of any of these three properties essentially leads to the same result in terms of protocol parameters. 

\section{Discussion}

\subsection{The standard solid-state protocols}

Based on the previous results, we define the standard solid-state protocols (SSSP) for plane-wave DFT calculations. As mentioned earlier, the name was chosen to mirror the standard solid-state pseudopotential libraries \cite{sssp}, indicating their seamless interoperability and the fact that they are designed to be used together.
%\todo{make sure how we want to approach this}.
Similar to Ref. \cite{sssp}, our benchmarks were done using the {\sc Quantum ESPRESSO} distribution, but users from other plane-wave pseudopotential codes may also benefit from these results (e.g. VASP \cite{kresseEfficientIterative1996, kresseEfficiencyAbinitio1996}, Abinit \cite{abinit}), as well as all-electron codes (e.g. Fleur \cite{fleurCode}, WIEN2k \cite{blaha2020wien2k}) that rely on smearing for facilitating convergence of BZ sampling (see Ref. \cite{bosoni_how_2023} for a review of protocols across different codes). As mentioned before, the numerical values obtained with Marzari-Vanderbilt cold smearing in these protocols may be used as a first approximation for other smearing techniques (e.g. Methfessel-Paxton), as long as they have similar dependencies of the free energy with respect to $\sigma$.

%\todo{Are we sure the BZ sampling works well in the same way as QE here?}. \red{GP: yes, see our verification in the Nat Rev Phys 2024. We should say "that rely on smearing for facilitating convergence of BZ sampling" or something similar. Not all do this - Green function methods, tetrahedra methods, ...}.

The protocols are designed to represent three distinct choices according to different precision and efficiency tradeoffs, and are named Fast, Balanced and Stringent. The values of the optimized parameters for each protocol can be directly read from the optimized Pareto front from Fig. \ref{fig:pareto_front_constrained} (indicated by black rectangles),
%\red{GP: can these 3 somehow be shown in the figure?},
and are summarized in Table \ref{tab:parameters}. In the following, we present a general overview of each protocol, along with their suggested SSSP pseudopotential family \cite{sssp}. We note that these choices are made based on the results obtained for metallic materials not including lanthanides nor actinides, and we suggest ways for which they can be adapted for insulators and metals with lanthanides/actinides.

\begin{table}[h!]
\centering
\begin{tabular}{m{7em} m{6em} m{6em} m{5em} m{17em}}
\toprule
 Protocol & $\sigma_{\textrm{cold}}$ (Ry) & $\sigma_{\textrm{cold}}$ (eV) &  $\lambda$ (\AA$^{-1}$) & Notes \\ 
\midrule
Fast & 0.0275 & 0.3742 & 0.30 & For testing purposes\\
\midrule
Balanced & 0.0200 & 0.2721 & 0.15 & Always recommended for insulators\\
\midrule
\multirow{2}{*}{Stringent} & \multirow{2}{*}{0.0125} & \multirow{2}{*}{0.1701} & \multirow{2}{*}{0.10} &  Always recommended for metals \\
 & & & & with lanthanides/actinides \\
\bottomrule
\end{tabular}

\caption{\textbf{Standard solid-state protocols and their respective optimized parameters}. $\sigma_{\textrm{cold}}$ represents the numerical values of smearing temperature found through the optimization using Marzari-Vanderbilt-DeVita-Payne cold smearing \cite{marzari1999thermal}, expressed both in Ry and eV (we note that, in the input of {\sc Quantum ESPRESSO}, the value is expressed in Ry).
%also valid for Methfessel-Paxton \cite{methfesselHighprecisionSampling1989} and Gaussian broadening functions\red{GP: Is this true???}. $\sigma_{\textrm{F.D.}}$ represents the analogous value for the smearing temperature using Fermi-Dirac broadening, scaled by $(\sqrt{2/3}\pi)^{-1}$ such as to match Gaussian total free energies (as prescribed by ref.\cite{dossantosFermiEnergy2023}).
$\lambda$ represents the optimized $k$-point distance for each choice of smearing, as defined in the \textit{Benchmarking workflows} section in the Results. As discussed in the main text, the Balanced protocol is always the optimal choice for insulators. The Stringent protocol, on the other hand, is the default recommendation for metallic systems containing lanthanide or actinide elements.
}
\label{tab:parameters}
\end{table}

\begin{itemize}[leftmargin=*, label={}]
    \item \textbf{Fast:} The fastest choice, meant for testing purposes or preliminary structure relaxations. It is not guaranteed to produce precise quantities with respect to the baseline reference, but it should still efficiently generate consistent and reproducible results. To be used alongside SSSP Efficiency pseudopotentials.

    \item \textbf{Balanced:} On average, 3.6x slower for metals than the Fast protocol. Suitable for high-throughput screening and most practical DFT applications. For metals, it yields an average error on forces of 20.9 meV/{\AA} and 4.4 meV/atom for total energies. The SSSP Efficiency pseudopotential family is the recommended choice in this case.

    \item \textbf{Stringent:} 2.7x slower than Balanced (or an order of magnitude slower than Fast) on average. Produces average errors on forces of 11.5 meV/{\AA } for 2D and 3D metals, and average errors on total energies of the order of 1.7 meV/atom. It is the highest precision choice before reaching the high slopes %\red{GP: Is it exponential? I think it just diverges - I guess it depends on how 'exponential' is defined, anyways}
    of the curve between required wall time $\times$ average errors. SSSP Precision pseudopotentials, along with their optimized plane-wave cutoff energies \cite{sssp} are suggested to be used here for maximum precision. 
\end{itemize}

% \begin{itemize}[leftmargin=*, label={}]
%     \item \textbf{Stringent:} Produces average errors on forces of 11.5 meV/{\AA } for 2D and 3D metals, and average errors on total energies of the order of 1.7 meV/atom. It is the best choice before reaching the ``exponential wall" on the curve between required wall time $\times$ average errors. The Precision class of pseudopotentials from SSSP, along with their optimized plane-wave cutoff energies \cite{sssp} are suggested to be used here, for maximum precision. 

%     \item \textbf{Balanced:} On average, 2.70x faster for metals than the Stringent protocol, being suitable for high-throughput screening. For metals, it yields an average error on forces of 20.9 meV/\AA, and 4.4 meV/atom for total energies. The SSSP Efficiency pseudopotential family is the recommended choice in this case.

%     \item \textbf{Fast:} 3.6x faster than Balanced, meant for mostly testing purposes or coarse structure relaxations. Not guaranteed to produce reliable results (but likely to do so), and calculations will still run without crashing errors. To be always used alongside SSSP Efficiency pseudopotentials.
% \end{itemize}

\subsection{Insulators and metals with open $f$-shell electrons}

The optimized parameters in the previous section are chosen based on an analysis of metals without lanthanides/actinides. However, other classes of materials may have different needs for parameters that optimize precision and efficiency. More specifically, insulators suffer much less from sampling errors and allow coarser meshes to be used, decreasing the computational cost required. On the other hand, materials with lanthanide or actinide elements particularly benefit from smaller smearing values due to high systematic errors, as discussed in the Results section.
%\red{GP: NOTE: in npj sections do NOT have numbers - so we need to give short clear names to sections, and refer to sections by name. Check everywhere}.
In this work, however, it has been decided that each protocol should represent a single and unified choice of parameters across material classes. This design choice is aimed at the simplicity of how each protocol is defined and is especially useful when the metallic/insulating nature of the material is not known a priori. In this way, we aim to solve the aforementioned issue by the following procedure:

$i$) \textbf{If the structure is insulating:} The Balanced protocol is the most recommended choice since it already provides average errors on forces that are one order of magnitude smaller than for metals. Running Stringent, in this case, would imply a cost 1.82x higher and only minor improvements in errors.

$ii$) \textbf{For metals including lanthanide/actinide elements:} Employing the Stringent protocol is always encouraged in this case. The Balanced protocol smearing value of 0.02 Ry yields higher errors on forces that lie above the default thresholds of {\sc Quantum ESPRESSO}.
% \begin{itemize}[leftmargin=*, label={}]
%     \item \textbf{If structure is insulating:} The Balanced protocol is the most recommended choice, once it already provides average errors on forces that are one order of magnitude smaller than for metals. Running Stringent, in this case, would imply a cost 1.82x higher and only minor improvements in errors.
%     \item \textbf{If metal with lanthanide/actinide elements:} Employing the Stringent protocol is always be encouraged in this case. The efficiency protocol's smearing value of 0.02 Ry yields higher errors on forces that lie above the default thresholds from {\sc Quantum ESPRESSO}.
% \end{itemize}

\begin{figure}[!h]
\centering
    { \hspace{0.6cm}
   \includegraphics[width=0.81\linewidth]{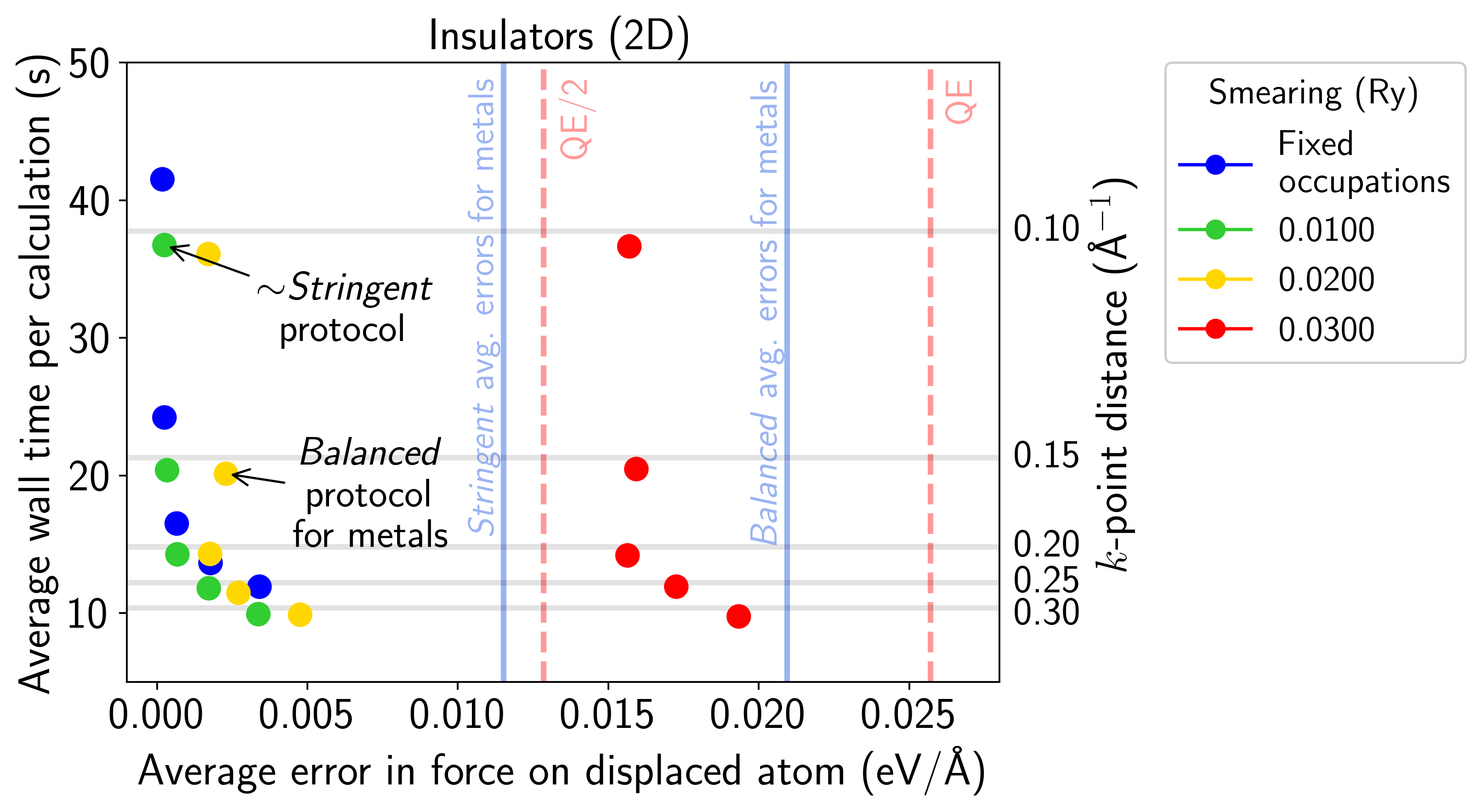}
   }
   \includegraphics[width=0.79\linewidth]{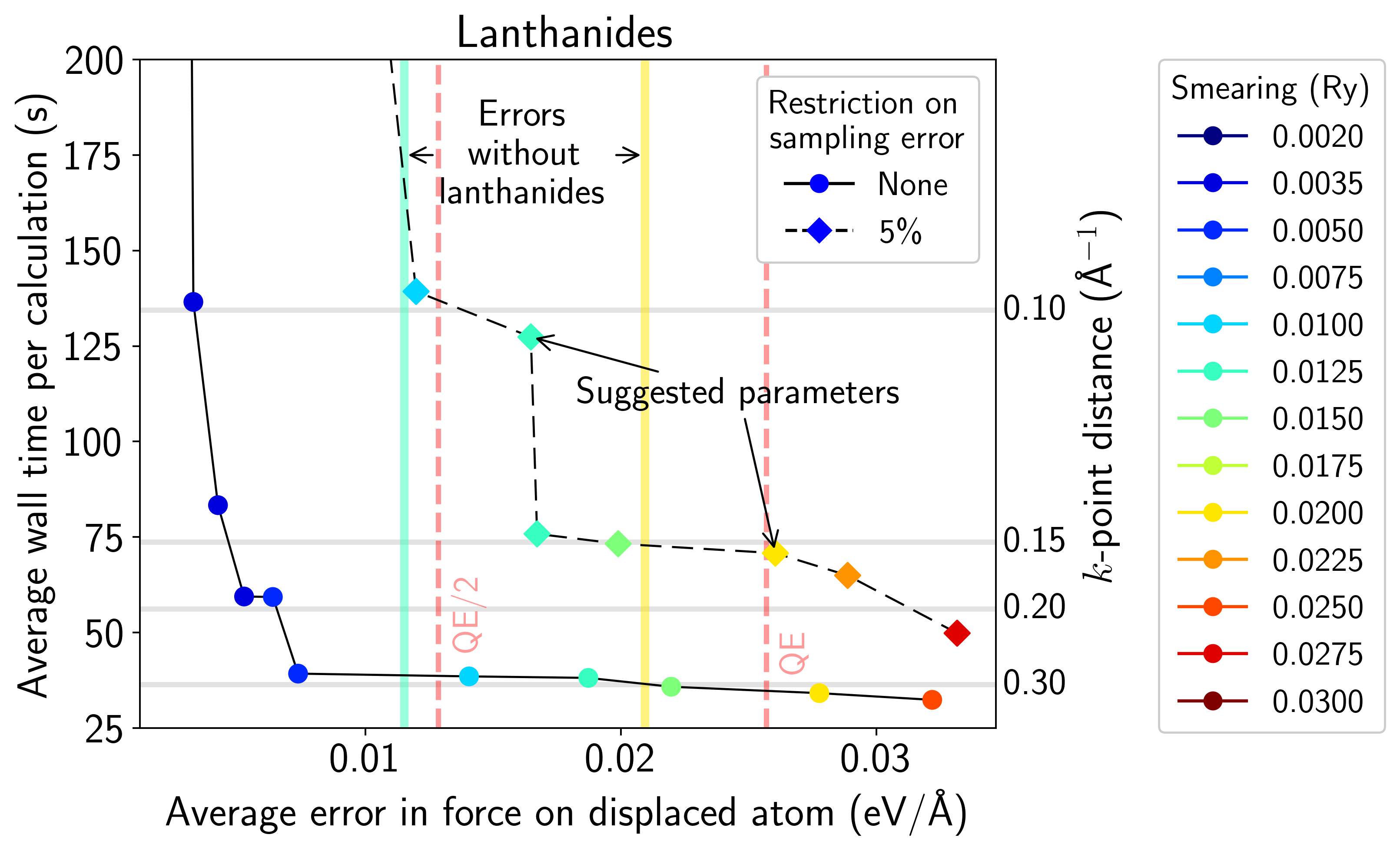}
    \caption{\textbf{Average wall time per calculation versus the average errors on the force on a displaced atom for insulators (top) and metals with lanthanides (bottom).} The parameters chosen for the protocols (Table \ref{tab:parameters}) are indicated in the plots by arrows. Additional vertical bars are also used to indicate the reference of average errors produced by the protocols on simple metals (Figure \ref{fig:pareto_front_constrained}). The term $\sim$Stringent protocol in Fig.(a) carries a $\sim$ symbol to indicate that it represents the parameters that are closest to the ones chosen for this protocol (0.0125 Ry instead of 0.01 Ry of smearing as calculated for insulators here). From the top panel, it is observed that the Stringent protocol is never needed for insulators, as the Balanced choice produces more efficient calculations with comparable errors. On the other hand, the bottom panel illustrates the need for the Stringent protocol for metals with lanthanides, in which the Balanced protocol produces excessive errors in these cases. Refer to Fig. \ref{fig:pareto_front_constrained} for further explanation of the plots.}
    \label{fig:insulators_and_lanthanides}
\end{figure}

These guidelines are justified in Figure \ref{fig:insulators_and_lanthanides}, where we show that running the Balanced protocol for insulators already produces satisfiable results with lower resource needs, but might be not enough for lanthanide materials, which can strongly benefit from the Stringent protocol. In cases where structures run in high-throughput workflows are unknown to be metallic or insulating, a standard practice can be to run non-lanthanide-containing structures with the Balanced protocol, then check their electronic structure and rerun it with the Stringent protocol for metallic cases, depending on the precision requirements.

\subsection{Benchmarks on ``higher-order" properties}

As a further benchmark and validation of the choice of protocol parameters we investigate, for a smaller subset of approximately 20\% of the metallic materials, how errors on total energies, forces and stresses propagate to other material properties, namely the equilibrium lattice constant and phonon frequencies. 
From our design choice of protocols, we can safely assume that the largest part of sampling errors are suppressed. This enables us to use smearing temperature as the unique variable for measuring errors. In this way, we compute the relaxed structure and $\Gamma$-phonon optical frequencies using only the densest $k$-point mesh used in this work (see Supplementary Table S1) and track how these values evolve as a function of $\sigma$. Supplementary Figure S4 shows the relative errors for relaxed lattice constants and optical phonon frequencies as a function of smearing temperature for the calculations of both 2D and 3D materials. From these results, we can estimate the average values of errors in these properties that our given choice of protocol parameters are expected to have, as displayed in Table \ref{tab:high-order-errors}.

\begin{table}
\centering
  \begin{tabular}{m{8em}m{5em}m{5em}m{5em}m{5em}}
    \toprule
    \multicolumn{1}{c}{} & \multicolumn{4}{c}{Average percentage error} \\
    \multicolumn{1}{c}{} & \multicolumn{2}{c}{Lattice constants} & \multicolumn{2}{c}{Largest freq. at $\Gamma$} \\
    \midrule
    Protocol & 2D & 3D & 2D & 3D \\
    \midrule
    Stringent & 0.04\% & 0.03\% & 0.26\% & 0.39\% \\
    Balanced & 0.1\% & 0.05\% & 0.41\% & 0.76\% \\
    \bottomrule
  \end{tabular}
\caption{\textbf{Estimate of average percentage errors for the Stringent and Balanced protocols}. Estimated errors are reported for equilibrium lattice constants and for the phonon frequency at $\Gamma$ of the highest (optical) branch.}
\label{tab:high-order-errors}
\end{table}

\subsection{Protocols availability}

Our protocols have been seamlessly integrated within the {\sc Quantum} {\sc ESPRESSO} environment and scaled to high-throughput calculations through the AiiDA framework, suiting different use cases. First, they are implemented within the Materials Cloud \cite{talirz2020materials} tool ``{\sc Quantum ESPRESSO} input generator".
%and structure visualizer'' \cite{mc-input-gen} \red{GP: maybe we can just remove "and structure visualizer" from here?}.
On the tool website, users can upload crystal structures in various formats and get input files generated automatically with meaningful input parameters chosen based on the benchmarks from this work, along with the SSSP pseudopotentials and related parameters extensively benchmarked by \citeauthor{sssp} \cite{sssp}. This provides a useful starting point for calculations with new structures and is especially helpful for beginners with the code. For automating this procedure, users of any workflow for the {\sc Quantum ESPRESSO} \texttt{pw.x} code in \texttt{aiida-quantumespresso} have access to these protocols and their parameters through the simple methods \texttt{get\_builder\_from\_protocol()} available in each of the AiiDA \texttt{WorkChain} workflow classes \cite{aiida-qe}. This also makes it suitable for running high-throughput workflows. Furthermore, for users outside the python ecosystem, our protocols are available in the AiiDAlab {\sc Quantum ESPRESSO} app \cite{qe-app}, which is an interactive platform with a GUI application where users can use ready-to-run workflows for materials properties and analyze results within a web browser. Being heavily automated and powered by AiiDA in the backend, these workflows can rely on our protocols as a part of setting up the required calculations on the fly.

% \red{GP: shall we turn the order saying that: 1. there is the QE input gen; 2. for any user wanting to automate this, tehre is the QE plugin; 2bis. this is suitable for high-throughput; 3. For those not wanting to use python scripts or the terminal, there is AiiDAlab, that internally uses the workflow from the QE plugin and exposes an easy GUI.}

\section{Conclusions}
This work identifies protocols for high-throughput DFT calculations by optimizing the precision $\times$ cost trade-off from the interplay between $k$-point sampling and smearing techniques. For a rigorous control of the sources of errors, we introduce an analysis that decouples numerical errors arising from smearing (systematic errors) and sampling errors due to finite $k$-point meshes. We then propose a criterion for suppressing the less controllable errors from $k$-sampling in favor of systematic errors, in a way that a reliable balance between numerical precision $\times$ computational cost can be achieved. By extensively benchmarking these parameters for a wide range of materials, we determine different choices of $k$-sampling and smearing parameters that best optimize the required computational cost for a given level of precision required. We then propose heuristics on how to apply those choices for different classes of materials, namely: metallic systems with lanthanide elements, in which systematic errors are chronically more severe, and insulators, that do not suffer from sampling errors as strongly. 
% \red{GP: I think we should have a paragraph here summarizing the main points of the paper in a more general way, e.g. we formally define 2 sources of errors, quantify their effect on materials properties. Only the outcome are the protocols, I would not start from those as the main point.}
% This work contributes to frameworks of high-throughput DFT calculations by introducing the concept of protocols. 
% By being extensively tested for a wide range of materials, protocols can reliably achieve different levels of precision and computational efficiency through their optimized parameters,

From these results, we define the standard solid-state protocols (SSSP), that are not only suitable for high-throughput workflows, but also for running new calculations for specific materials with minimal initial investigation. By automatically providing input parameters for any structure, these protocols act as abstractions, hiding layers of complexity of the code's internal algorithms. This is beneficial for making the code both more robust and user-friendly, avoiding potential errors due to inadequate parameter selection. More generally, protocols can be an essential piece in building interoperable software suitable for robust computational workflows. They enable workflows to be treated as building blocks of larger workflows for materials design, suiting different needs and requirements. 

Here, we specifically establish protocols for self-consistent DFT calculations in {\sc Quantum ESPRESSO} by benchmarking the most fundamental quantities of the theory, i.e. total energies, forces, and stresses. In this way, our protocols provide the basic parameters related to Brillouin zone sampling, in addition to pseudopotentials and plane-wave cutoffs already provided by Ref. \cite{sssp}. 
We note, though, that further tests may be required when running calculations targeted at other properties, where also additional parameters may need to be considered. For instance, in the computation of linear-response properties such as phonons, convergence of the $\mathbf{q}$-grid of monochromatic perturbations is also needed. In these cases, we also note the emergence of advanced Brillouin zone integration methods using nonuniform grids \cite{ChenEtAlNonuniformGridsBrillouin2022} as promising approaches. Finally, similar techniques as those developed in this work can be applied to create protocols for many other simulation codes beyond {\sc Quantum ESPRESSO}, either by direct application of the numerical results of this work, when applicable, or by rerunning the benchmark, e.g. if the smearing or $k$-sampling stategies are different. Importantly, we note that our results are expected to be independent of the basis set used (plane waves, Gaussians, localized orbitals) or the all-electron vs. pseudopotential implementation (norm conserving, ultrasoft, projector-augmented waves) and thus broadly universal.
%By identifying the free parameters that arise from implementation details of the theory and benchmarking them through a wide spectrum of materials, the main goal becomes translating the domain-specific knowledge of scientific codes into an interoperable piece of software that can benefit a larger community within materials modeling. 
% \red{GP: this last sentence seems very generic and unclear. I would remove the last part, and just say that the results are valid beyond QE (in some cases even quantitatively).}

\section{Methods}
\label{sec:methods}

The crystal structures in this work were taken from the Materials Cloud three-dimensional and two-dimensional crystals databases (MC3D and MC2D) \cite{mounetTwodimensionalMaterials2018b, campiExpansionMaterialsCloud2023, MC2D, MC3D}. These are open materials databases of computational DFT simulations of crystals originating from experimentally known compounds from MPDS \cite{mpds1}, COD \cite{cod}, and ICSD \cite{icsd}. The compounds used in our calculations were randomly selected among the structures containing at most 14 atoms in the unit cell available in the databases. Supplementary Table S2 displays the number of structures selected for the DFT benchmarking workflows. The exact material structures used, along with their calculation results, are available in the Materials Cloud Archive enry associated to this work \cite{MCArchiveDraft} in the form of AiiDA archive files.

% \red{GP: Say also that there is a Materials Cloud Archive entry with the exact materials chosen, as well as all calcualtion results, and the AiiDA archive file. Create already a draft so you get a DOI and can already cite the entry, even if it's not public yet.}

The computational workflows were developed and managed with AiiDA \cite{aiida, aiida1, uhrin2021workflows}, providing seamless scalability and provenance tracking of all the calculation results. The DFT calculations were run with {\sc Quantum ESPRESSO} v7.0 \cite{giannozzi2009quantum, giannozzi2017advanced, giannozzi2020quantum}, employing the standard solid-state pseudopotentials family SSSP PBEsol Efficiency 1.2 \cite{sssp} and PBEsol as exchange-correlation functional \cite{pbesol}. Thanks to the AiiDA plugin for {\sc Quantum ESPRESSO} \cite{aiida-qe}, inputs for calculations were automatically generated with the appropriate recommended pseudopotentials and the corresponding
%pseudopotentials for each structure and their respective recommended 
plane-wave energy cutoffs, as extensively benchmarked by \citeauthor{sssp} \cite{sssp}. Furthermore, a stricter threshold of $0.2\cdot 10^{-9}$ Ry/atom for the self-consistent field (SCF) electronic minimization was employed. As the average number of atoms per structure was 5.2, this renders an average SCF threshold of approximately $10^{-9}$ Ry, being three orders of magnitude smaller than the one employed by default in {\sc Quantum ESPRESSO} ($10^{-6}$ Ry). The same parameters were employed for the relaxation calculations, using thresholds for ionic minimization of $0.5\cdot10^{-5}$ Ry/atom for total energies, $0.5\cdot10^{-5}$ Ry/$a_0$ for forces and 0.1 kbar for pressure. $\Gamma$-phonon frequencies were calculated from these geometries through density-functional perturbation theory (DFPT), as implemented in the \texttt{ph.x} module from {\sc Quantum ESPRESSO}. One assumption of this work is that the optimized parameters obtained from the analysis of these results are transferable to other calculations with similar settings, i.e., other exchange-correlation functionals (e.g., LDA or PBE) and pseudopotential families (e.g., SSSP Precision \cite{sssp}).

\section{Data Availability}

Full provenance of the high-throughput DFT calculations in this work is generated through AiiDA and available in the Materials Cloud Archive \cite{MCArchiveDraft}, along with a Jupyter Notebook for generating interactive plots of Figure \ref{fig:2D-3D-heatmap} with variable error thresholds. The input generator for {\sc Quantum Espresso} is available at \url{https://qeinputgenerator.materialscloud.io}. The high-throughput compatible implementation of the protocols are available in the AiiDA plugin for {\sc Quantum Espresso} (source code at \url{https://github.com/aiidateam/aiida-quantumespresso}).
%Interactive plots with the data of general convergence of materials are also available in \todo{decide where we will publish it}.

\section{Acknowledgements}

G.d.M.N. acknowledges financial support from École Polytechnique Fédérale de Lausanne, through its Master Excellence fellowships. 
% N.M. and G.d.M.N. \red{GP: All? At leats also me and Marnik and Flaviano} to the Swiss National Science Foundation (SNSF), through its National Centre of Competence in Research (NCCR) MARVEL 
% \todo{Clarify funding sources from everyone}.
% \red{GP: check \url{https://nccr-marvel.ch/miscellaneous/marvel-toolbox} for the exact string to use, include also the Grant number for phase 3}.
This research was supported by the NCCR MARVEL, a National Centre of Competence in Research, funded by the Swiss National Science Foundation (grant number 205602). % Exact string from MARVEL %\url{https://nccr-marvel.ch/miscellaneous/marvel-toolbox}
We acknowledge access to Alps at the Swiss National Supercomputing Centre, Switzerland under the MARVEL's share with the project ID mr32. % \url{https://www.cscs.ch/user-lab/code-of-conduct}
% We acknowledge access to computational resources from the Alps (Eiger) supercomputer at the Swiss National Supercomputing Centre (CSCS), with project ID mr32 \todo{check project}.
% \red{GP: check \url{https://www.cscs.ch/user-lab/code-of-conduct}, it's the last one "Contractual partners" for mr32. I guess we want to say "MARVEL share", and say it's Eiger on Alps indeed.}

\section{Author Contributions} 
 G.d.M.N. implemented the computational workflows and pipeline for analysis of the results and wrote the initial manuscript. F.J.d.S. contributed with fundamental discussions on the theory of smearing and Brillouin zone sampling. M.B. and D.G. provided technical support on the data source infrastructure. All authors participated in the discussion for the design of the methods of analysis and the criteria used for the creation of the protocols and contributed to writing the final manuscript. G.P. and N.M. supervised the project.

\bibliography{main}% Produces the bibliography via BibTeX.

\end{document}

\title{Supplementary Information: \\
Accurate and efficient protocols for high-throughput first-principles materials simulations}% Force line breaks with \\

\author{Gabriel M. Nascimento}
\affiliation{Theory and Simulation of Materials (THEOS), École Polytechnique Fédérale de Lausanne, Lausanne 1015, Switzerland} 
\affiliation{PSI Center for Scientific Computing, Theory and Data, 5232 Villigen PSI, Switzerland} 
\affiliation{National Centre for Computational Design and Discovery of Novel Materials (MARVEL), Paul Scherrer Institute PSI, 5232 Villigen PSI, Switzerland} 

\author{Flaviano José dos Santos}
\author{Marnik Bercx}
\affiliation{PSI Center for Scientific Computing, Theory and Data, 5232 Villigen PSI, Switzerland} 
\affiliation{National Centre for Computational Design and Discovery of Novel Materials (MARVEL), Paul Scherrer Institute PSI, 5232 Villigen PSI, Switzerland} 

\author{Davide Grassano}
\affiliation{Theory and Simulation of Materials (THEOS), École Polytechnique Fédérale de Lausanne, Lausanne 1015, Switzerland} 

\author{Giovanni Pizzi}
\affiliation{PSI Center for Scientific Computing, Theory and Data, 5232 Villigen PSI, Switzerland} 
\affiliation{National Centre for Computational Design and Discovery of Novel Materials (MARVEL), Paul Scherrer Institute PSI, 5232 Villigen PSI, Switzerland} 

\author{Nicola Marzari}
\affiliation{Theory and Simulation of Materials (THEOS), École Polytechnique Fédérale de Lausanne, Lausanne 1015, Switzerland} 
\affiliation{PSI Center for Scientific Computing, Theory and Data, 5232 Villigen PSI, Switzerland} 
\affiliation{National Centre for Computational Design and Discovery of Novel Materials (MARVEL), Paul Scherrer Institute PSI, 5232 Villigen PSI, Switzerland} 

\maketitle

\newpage

\section{Supplementary Figures}

\begin{figure}[h]
    \centering
    \includegraphics[width=1\linewidth]{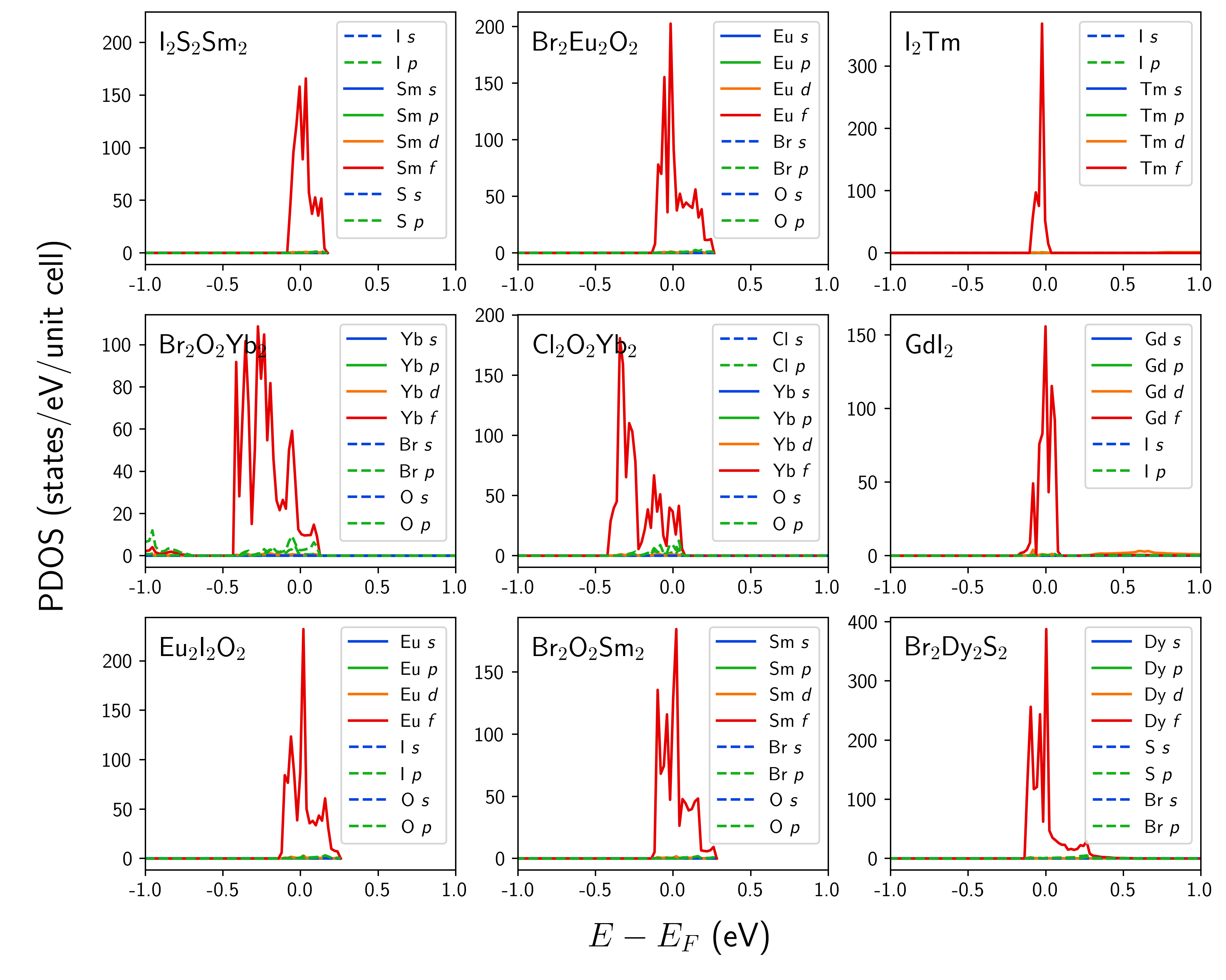}
        \caption{Projected density of states (PDOS) of materials containing lanthanides with the highest systematic errors observed in Fig. \ref{fig:systematic_errors_lanthanides} of the main paper. In every case, the presence of a high number of $f$-states introduced by the lanthanide element (red lines) around the Fermi level generates sharp peaks in the density of states of these materials.}
    \label{fig:PDOS_lanthanides}
\end{figure}

\begin{figure}[h]
    \centering
    \includegraphics[width=1\linewidth]{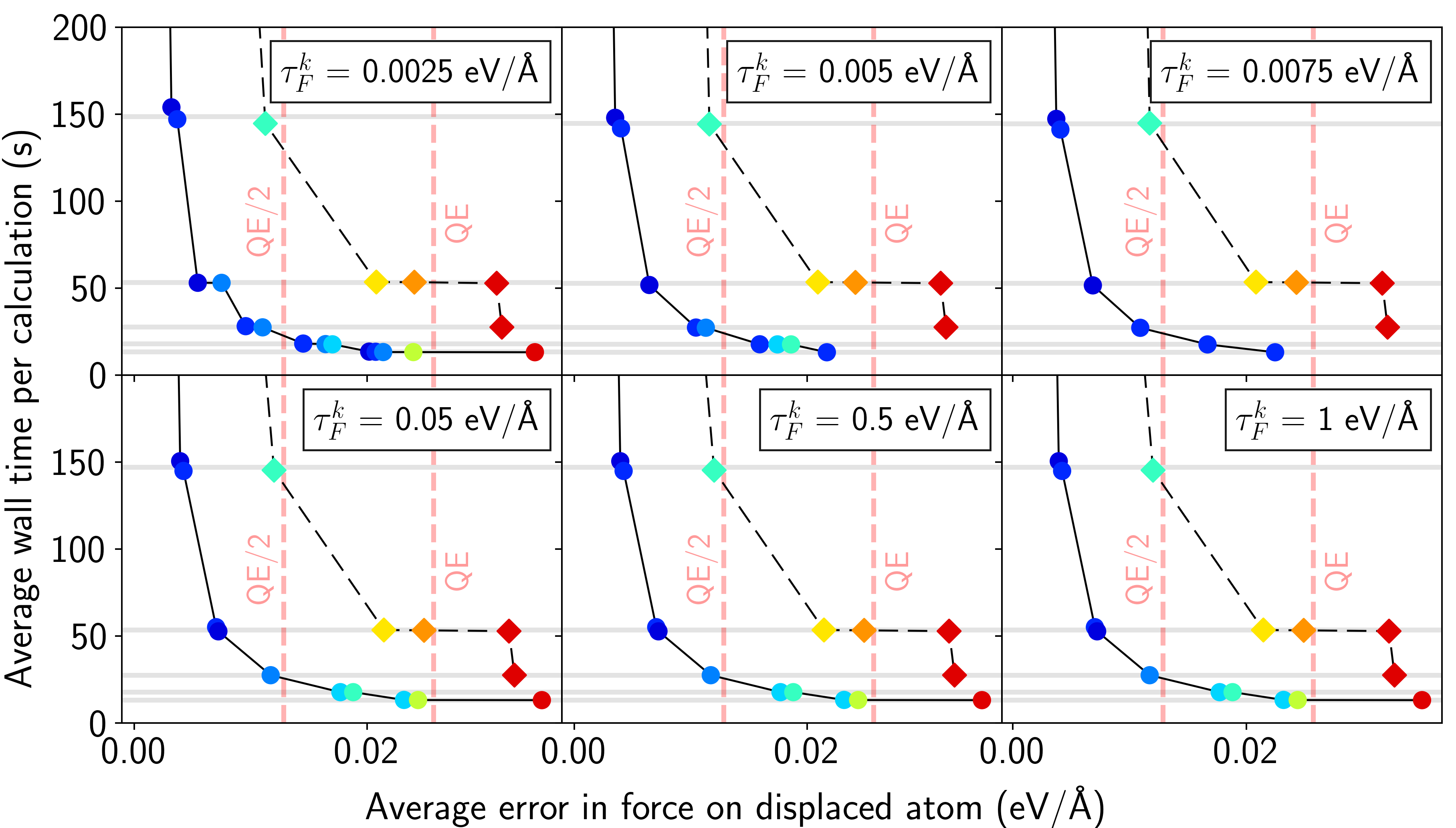}
    \caption{Pareto front analysis as presented in Figure \ref{fig:pareto_front_constrained} of the main paper, for different values of the threshold on sampling errors ($\tau^k$) used to determine $\sigma_\textrm{ref}$ of each material. As in Fig. \ref{fig:pareto_front_constrained}, the dashed lines represent the Pareto front with the additional constraint on the composition of the errors and are independent of the value of $\tau^k$ in this range, as opposed to the unconstrained version (continuous lines).}
    \label{fig:tk_analysis}
\end{figure}

\begin{figure}[h]
    \centering
    \includegraphics[width=0.7\linewidth]{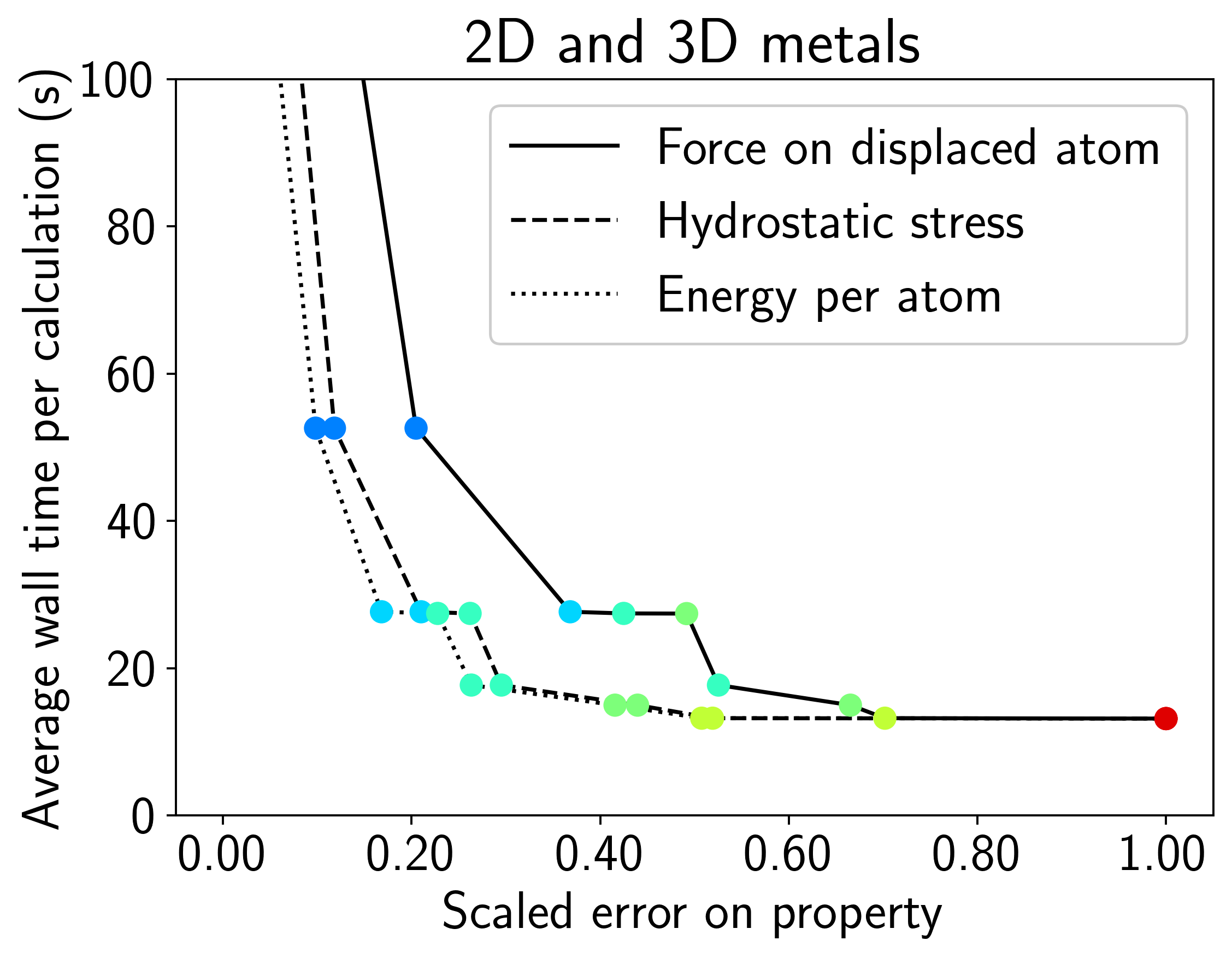}
    \caption{Pareto front for all three properties tracked in the benchmarking calculations.
    The $y$-axis displays the average computational wall time per calculation, while the $x$-axis represents a min-max normalized error on total energies, forces, and stresses. More precisely, as each combination of parameters $(\lambda, \sigma)$ produces an average total error on property $P$ of $\bar\varepsilon_P(\lambda, \sigma)$, the values plotted on the $x$-axis are defined to be $x = [\bar\varepsilon_P - \textrm{min}(\bar\varepsilon_P)]/[\textrm{max}(\bar\varepsilon_P) - \textrm{min}(\bar\varepsilon_P)]$, where $\textrm{min}(\bar\varepsilon_P)$ and $\textrm{max}(\bar\varepsilon_P)$ are the minimum and maximum value of $\bar\varepsilon_P$ over all possible $(\lambda, \sigma)$ combinations calculated in this work. As in other plots, each point represents a pair of parameters and is colored according to the value of $\sigma$. As mentioned before, the $y$ coordinate of each point has a direct mapping to its $\lambda$ value, such that the optimized parameters for each property can be directly inferred and are essentially the same.}
    % displaying the average computational wall time on the $y$-axis and the normalized error on total energies, forces, and stresses on the $x$-axis.  \todo{should it also be moved to the supplementary?}}
    \label{fig:property_comparison}
\end{figure}

\begin{figure}
    \centering
    \includegraphics[width=\linewidth]{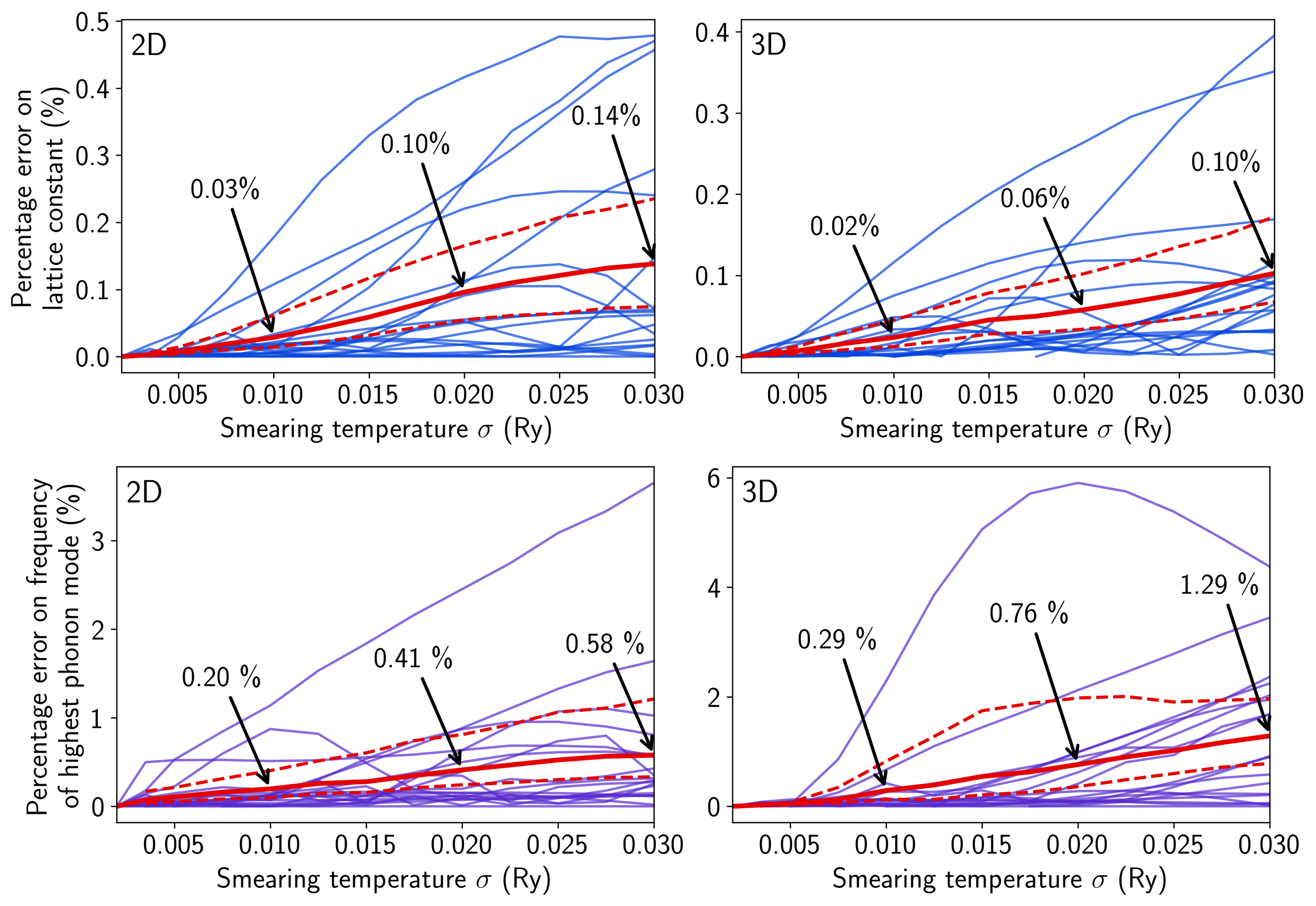}
    \caption{Relative errors on ``higher-order" properties as a function of smearing. Plots on the top: percentage errors on relaxed lattice constant for 2D (left) and 3D (right) metals. Plots on the bottom: percentage errors calculated phonon frequency of the highest-energy optical phonon mode at $\Gamma$. Continuous red lines represent averaged error values, with black arrows displaying their numerical values at $\sigma= 0.1,\, 0.2,\, 0.3$ Ry. Dashed lines indicate estimates for the 95\% confidence intervals of the average error, calculated through bootstrapping. In all cases, calculations are carried out using the densest $k$-point meshes, and the smearing reference is taken according to $\sigma_{\textrm{ref}}$ defined for each material (see section on analysis of errors in the main paper).} 
    \label{fig:lattice_constant_and_phonon_erros}
\end{figure}

\newpage

\section{Supplementary Tables}

\begin{table*}[h]
    \centering
    \begin{tabular}{lcc}
    \toprule
         & Smearing temperature $\sigma$ (Ry) & $k$-point distance $\lambda$ (1/\AA) \\
        \midrule
        Metals & 0.002, 0.0035, (0.005 - 0.030, 11) & \multirow{2}{*}{ 0.035, (0.05 - 0.30, 6)} \\
        Insulators &   0.0*, 0.01, 0.02, 0.03  &  \\
     \bottomrule
    \end{tabular}
    \caption{Smearing temperature values ($\sigma$) and $k$-point sampling distances ($\lambda$) employed in the benchmarking of DFT calculations in the workflows. The notation ($x$ - $y$, $n$) represents a equally spaced sequence of $n$ values within $x$ and $y$, both included. The value 0.0* represents DFT calculations run with smearing disabled (fixed occupations).}
    \label{tab:workflow_parameters}
\end{table*}

\begin{table}[h]
    \centering
    \begin{tabular}{clc}
        \toprule
        \multicolumn{2}{c}{Class} & \# structures \\
        \midrule
        \multirow{3}{*}{2D} & insulators & 45 \\
         & metals without lanthanides & 59 \\
         & metals with lanthanides & 20 \\
        \midrule
        % \multirow{3}{*}{3D} &  &  \\
         3D & metals & 145 \\
         \bottomrule
    \end{tabular}
    \caption{Number of structures selected for the DFT benchmarks, separated by material class.}
    \label{tab:1-structures}
\end{table}